\documentclass[useAMS,usenatbib]{mn2e}
\usepackage{amsmath, amssymb}
\usepackage{graphicx}
\usepackage{wasysym}

\title[Modelling element abundances]{Modelling Element Abundances in Semi-analytic Models of Galaxy Formation}
\author[Yates et al.]{Robert M. Yates$^{1}$\thanks{Email: robyates@mpa-garching.mpg.de}, Bruno Henriques$^{1}$, Peter A. Thomas$^{2}$, Guinevere Kauffmann$^{1}$,\and Jonas Johansson$^{1}$ \& Simon D. M. White$^{1}$\\\\
$^{1}$ Max Planck Institut f$\ddot{u}$r Astrophysik, Karl-Schwarzschild-Str. 1, 85741, Garching, Germany\\
$^{2}$ Astronomy Centre, University of Sussex, Falmer, Brighton BN1 9QH, UK}

\begin{document}
\date{Accepted ??. Received ??; in original form ??}
\maketitle

\begin{abstract}
We update the treatment of chemical evolution in the Munich semi-analytic model, \textsc{L-Galaxies}. Our new implementation includes delayed enrichment from stellar winds, supernov\ae{} type II (SNe-II) and supernov\ae{} type Ia (SNe-Ia), as well as metallicity-dependent yields and a reformulation of the associated supernova feedback. Two different sets of SN-II yields and three different SN-Ia delay-time distributions (DTDs) are considered, and eleven heavy elements (including O, Mg and Fe) are self-consistently tracked. We compare the results of this new implementation with data on a) local, star-forming galaxies, b) Milky Way disc G dwarfs, and c) local, elliptical galaxies. We find that the $z=0$ gas-phase mass-metallicity relation is very well reproduced for all forms of DTD considered, as is the [Fe/H] distribution in the Milky Way disc. The [O/Fe] distribution in the Milky Way disc is best reproduced when using a DTD with $\leq50$ per cent of SNe-Ia exploding within $\sim 400$ Myrs. Positive slopes in the mass-[$\alpha$/Fe] relations of local ellipticals are also obtained when using a DTD with such a minor `prompt' component. Alternatively, metal-rich winds that drive light $\alpha$ elements directly out into the circumgalactic medium also produce positive slopes for all forms of DTD and SN-II yields considered. Overall, we find that the best model for matching the wide range of observational data considered here should include a power-law SN-Ia DTD, SN-II yields that take account of prior mass loss through stellar winds, and some direct ejection of light $\alpha$ elements out of galaxies.
\end{abstract}


\section{Introduction} \label{sec:Introduction}
\LARGE{S}\normalsize ignificant progress has been made in the field of galactic chemical evolution (GCE) since the first postulation of stellar nucleosynthesis by Arthur Eddington in the 1920s \citep{E20}. The first techniques to determine element abundances in both gas (e.g. \citealt{A42}) and stars (e.g. \citealt{CA51}) were developed, and the theory of stellar nucleosynthesis was given a more formal footing by \citet{BBFH57}. Later, more sophisticated studies of GCE were stimulated by the celebrated review by Beatrice Tinsley \citep{T80}. Now, it has been determined that a galaxy's metallicity is related to its luminosity (e.g. \citealt{L79}), age (e.g. \citealt{E93}, but see e.g. \citealt{F95}), and stellar mass (e.g. \citealt{T04}), and that different types of stars contribute to GCE in different ways (e.g. \citealt{SC05}).

However, many questions relating to the cosmic abundances of heavy elements still remain. For example, it is still unclear what exact role different types of supernov\ae{} (SNe) and stellar winds play in the chemical enrichment of galaxies (e.g. \citealt{McW97}), what the shape and universality of the stellar initial mass function (IMF) is (e.g. \citealt{BCM10}), how best to model the metal yields produced in stars (e.g. \citealt{R10}), and what the progenitors and delay times of SNe-Ia are (e.g. \citealt{MM12}). These are important questions for us to address, as the chemical evolution of galaxies plays a key part in the evolution of galaxies in general; the presence of metals affects the cooling of gas (e.g. \citealt{SD93}), the formation of stars (e.g. \citealt{W11}), stellar evolution (e.g. \citealt{SC05}), and the yields of newly synthesised metals (e.g. \citealt{WW95}) which are released into the interstellar medium (ISM), circumgalactic medium (CGM) and even the intergalactic medium (IGM).

Aside from the ongoing observational studies into these questions, galaxy evolution models incorporating sophisticated GCE modelling also provide an opportunity to further constrain the chemical evolution of galaxies. Many previous works have focused on reproducing the chemical signitures found in the solar neighbourhood (e.g. \citealt{T80,MG86,M86,TGB98,F04,DR09,CM09,C10,R10,TWS12,P12,C12}), chiefly in order to constrain the contributions from different types of SNe and stellar winds. Many others have focused on the chemical properties of local elliptical galaxies (e.g. \citealt{M94,TK99,TGB99,PM04,N05b,P09a,P09b,CM09,A10a,CM11,PM11}), chiefly to try to reconcile the observed positive slope in the relation between stellar mass ($M_{*}$) and $\alpha$ enhancement ([$\alpha$/Fe]) with our theoretical understanding of metal production and galaxy formation.

The aim of this work is to address both of these issues, using a new implementation of detailed chemical enrichment in the Munich semi-analytic model of galaxy formation. We investigate if the chemical properties of Milky Way (MW) disc stars and local elliptical galaxies can be simultaneously obtained with a self-consistent model which assumes a $\Lambda$CDM hierarchical merging scenario and varied star formation histories (SFHs). We also compare different SNe-II yield sets and SN-Ia delay-time distributions (DTDs), to see which allow us to best match the observational data considered.

This paper is structured as follows: in \S \ref{sec:L-Galaxies} we give a general outline of the Munich semi-analytic model, \textsc{L-Galaxies}. In \S \ref{sec:GCE ingredients} we describe the stellar yields, lifetimes and IMF used as inputs to our model. In \S \ref{sec:The GCE Equation} we describe the basic equations required to model GCE and discuss the SN-Ia DTD. In \S \ref{sec:Implementation} we explain how this GCE model is implemented into the larger semi-analytic model and review the key physical processes governing the distribution of metals throughout galaxies. In \S \ref{sec:General Results} we discuss our model results for the chemical composition of a) local, star-forming galaxies, b) the G dwarfs of the MW disc, and c) the stellar components of local ellipticals, and compare these results to the latest observations. We conclude our work in \S \ref{sec:Conclusions}.

\section{The semi-analytic model} \label{sec:L-Galaxies}
\textsc{L-Galaxies} \citep{S01,DL04,S05,C06,DLB07,G10,G13,H13} is a semi-analytic model of galaxy evolution which extends the methods set out in \citet{WF91,KWG93,K99} so that the model can be run on subhalo trees built from DM N-body simulations such as the \textsc{Millennium} \citep{S05}. Galaxy evolution is governed by the transfer of mass among the various components of a galaxy (disc stars, bulge stars, halo stars, cold gas, hot gas, central black hole, and ejecta reservoir), according to physical laws motivated by observations and simulations. \textsc{L-Galaxies} is currently able to reproduce the large-scale clustering of galaxies, the Tully-Fisher relation, and the optical colours, stellar mass function and gas-phase mass-metallicity relation observed in the local Universe. The model can also reproduce the abundance of galaxies as a function of stellar mass or luminosity out to $z=3$. Analytical treatments of gas stripping and tidal disruption of satellites, as well as SN and AGN feedback are included (see \citealt{G10}). The processes already included in \textsc{L-Galaxies} that are of most relevance to this work are reviewed briefly in \S \ref{sec:Infall, Cooling and Outflows}.

Prior to this work, \textsc{L-Galaxies} included a simple GCE implementation. A fixed metal yield of $0.03\cdot{}\Delta M_{*}$ was assumed to be ejected into the ISM immediately after a star formation event, where $\Delta M_{*}$ is the mass of stars formed at that time. A further 40 per cent of $\Delta M_{*}$ was assumed to return immediately to the gas phase as H and He. Such an `instantaneous recycling approximation' is often used in galaxy formation models for its simplicity, but does not adequately describe the delayed enrichment of metals, particularly from long-lived low- and intermediate-mass stars and SNe-Ia. Previously, \textsc{L-Galaxies} also did not consider individual chemical elements, but instead tracked only the total metal mass in each galaxy component. The tracking of individual elements allows us to compare with more detailed observational data on the chemical composition of the Milky Way and other galaxies (see \S \ref{sec:General Results}). For example, the ratio of $\alpha$ elements to iron is believed to be a good indicator of the star formation timescale. A comparison of [$\alpha$/Fe] between real galaxies and model galaxies with known star formation histories will allow us to test this. Also, in future, tracking individual elements will provide a more realistic treatment of gas cooling, which depends not only on the total metallicity, but also on the relative abundance of different heavy elements, as well as the ultraviolet background radiation.

The model parameters we use in this paper are identical to those in \citet{G10}, with the exception of the `halo-velocity-dependent SN energy efficiency', $\epsilon_{\textnormal{h}}$, which we have increased in order to maintain the same total SN feedback energy that was used previously (see \S \ref{sec:SN feedback}).

\section{GCE Ingredients} \label{sec:GCE ingredients}
In order to model the chemical evolution of galaxies, we first need to know the total mass of heavy elements liberated from stars at any given time. To do this, we need to know a) how many stars eject metals at that time, and b) how much of each element they eject. The former is given by the assumed stellar lifetimes, the IMF and the SFHs of galaxies. The latter is given by the stellar yields, obtained from stellar evolution models.

The yields, as well as depending on the initial mass (and metallicity) of the star, also depend on the mode of ejection. We consider three modes in this work; stellar winds from low- and intermediate-mass stars during their thermally-pulsing asymptotic giant branch phase (TP-AGB, or simply AGB phase), SNe-Ia from some intermediate-mass binary systems, and the SN-II explosions of massive stars. Each of these three modes releases a different set of heavy elements at different times. Long-lived stars of mass $0.85\lesssim M/\textnormal{M}_{\textnormal{\astrosun}} \lesssim 7$ release mainly He, C and N. SNe-Ia produce and eject mainly Fe and other iron-peak elements, whether they originate from single degenerate binaries \citep{WI73}, double degenerate binaries \citep{W84,IT84}, or otherwise (e.g. the binary progenitors of double-detonation, sub-Chandrasekhar-mass explosions, see \citealt{R11}). Finally, short-lived stars of mass $\gtrsim 7 \textnormal{M}_{\textnormal{\astrosun}}$ explode as core-collapse SNe-II, ejecting chiefly $\alpha$ elements (e.g. O, Ne, Mg, Si, S and Ca).

We note here that we only consider eleven chemical elements in our GCE model, namely, H, He, C, N, O, Ne, Mg, Si, S, Ca and Fe, as these elements are included in all of the yield sets we consider.

The following sub-sections outline in more detail these key ingredients for galactic chemical enrichment. The SFHs of galaxies are tracked self-consistently in our semi-analytic model and are discussed in \S \ref{sec:SFH, ZH and EH arrays}.

\subsection{The IMF} \label{sec:The IMF}
The IMF, $\phi(M)$, is a probability density function, which tells us the fraction of stars in a $1 \textnormal{M}_{\textnormal{\astrosun}}$ simple stellar population (SSP) that are within a given mass range. It is obtained from the observable present day mass function (PDMF) of field stars in the Milky Way, or from PDMF indicators in extragalactic regions. In this work, we assume that the IMF is the same in all regions of space and does not evolve with time. There are, however, currently conflicting conclusions in the literature as to its universality (e.g. \citealt{WK06}; \citealt{El06}; \citealt{BCM10}; \citealt{vDC10}; \citealt{G11}; \citealt{FDK11}; \citealt{CvD12b}).

The IMF used in this work is taken from \citet{C03}. This version is commonly used in chemical enrichment models, and is already utilised in \textsc{L-Galaxies} via the stellar population synthesis models of \citet{BC03,M05}. It's use therefore provides both a good comparison to other works and self-consistency within the code. The Chabrier IMF is given analytically as

\begin{align} \label{eqn:ChabrierIMF}
\phi(M)= \bigg{\{} & \begin{array}{ll}
													A_{\phi}M^{-1}e^{-(\textnormal{log }M-\textnormal{log }M_{c})^{2}/2\sigma^{2}} & \textnormal{if } M \leqslant 1 \textnormal{M}_{\textnormal{\astrosun}} \\
													B_{\phi}M^{-2.3} & \textnormal{if } M > 1 \textnormal{M}_{\textnormal{\astrosun}}
									 \end{array} \;\;,
\end{align}
where $M_{c}=0.079 \textnormal{M}_{\textnormal{\astrosun}}$ and $\sigma = 0.69$. The values of the coefficients $A_{\phi}$ and $B_{\phi}$ are determined by requiring that a) the overall function is continuous, and b) the IMF \textit{by mass} is normalised to $1 \textnormal{M}_{\textnormal{\astrosun}}$ over the full mass range of stars considered;

\begin{figure}
\centering
\includegraphics[totalheight=0.28\textheight, width=0.46\textwidth]{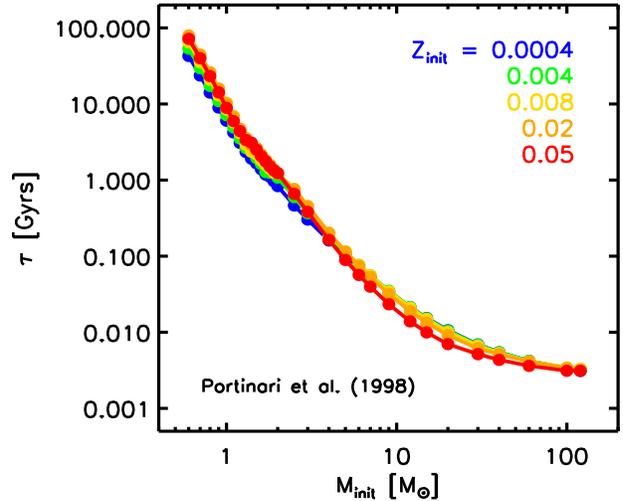}
\caption{The lifetimes of stars as a function of initial mass, for five different initial metallicities, as predicted by \citet{P98}.}
\label{fig:lifetimes}
\end{figure}

\begin{figure*}
\centering
\includegraphics[totalheight=0.46\textheight, width=0.71\textwidth]{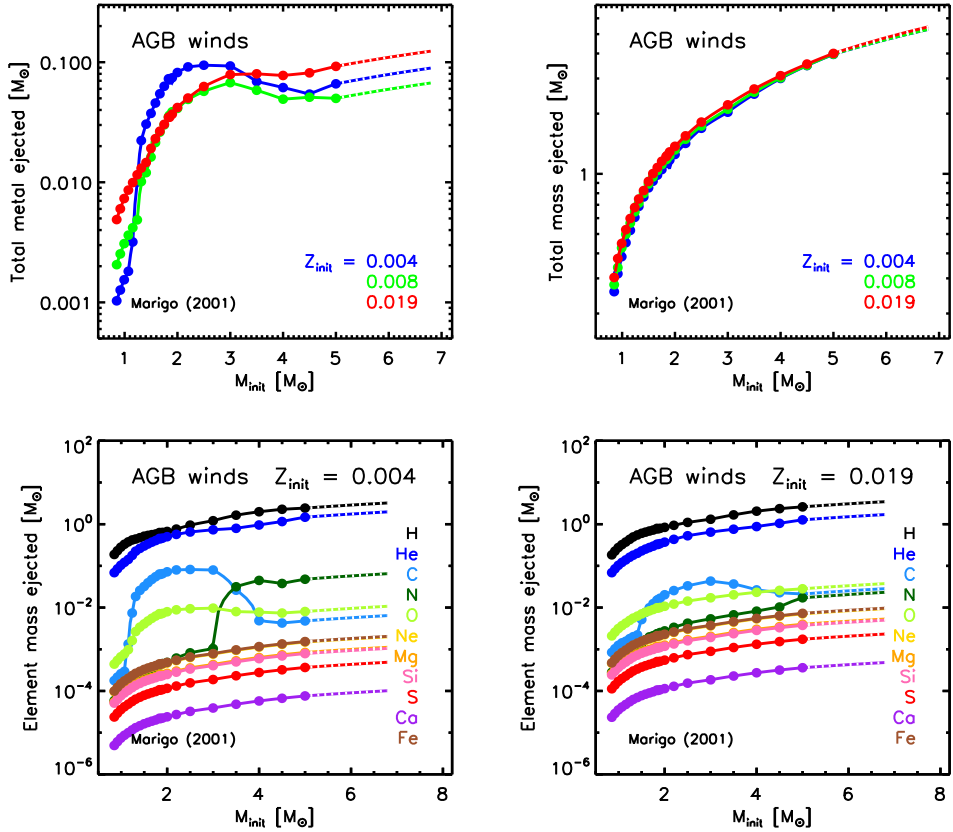}
\caption{Mass released by AGB winds from the \citet{M01} yield tables. Points indicate values from the yield tables. Solid lines indicate the interpolation used between these points. Dashed lines indicate extrapolations beyond the masses originally modelled. \textit{Top left}: The mass of metals ejected as a function of mass, for three different initial metallicities. \textit{Top right}: The total baryonic mass ejected as a function of mass, for three different initial metallicities. \textit{Bottom left}: The mass of each element ejected as a function of mass, for stars of $Z_{0}=0.004$. \textit{Bottom right}: Same as bottom left, for stars of $Z_{0}=0.019$.}
\label{fig:agbwinds}
\end{figure*}

\begin{equation} \label{eqn:IMFnorm}
\int^{M_{\textnormal{max}}}_{M_{\textnormal{min}}} M\phi(M) \textnormal{dM} = 1 \textnormal{M}_{\textnormal{\astrosun}} \;\;,
\end{equation}
where $M_{\textnormal{min}} = 0.1 \textnormal{M}_{\textnormal{\astrosun}}$. When assuming $M_{\textnormal{max}} = 120 \textnormal{M}_{\textnormal{\astrosun}}$, as we do in this work, the coefficients in Eqn. \ref{eqn:ChabrierIMF} are $A_{\phi}=0.842984$ and $B_{\phi}=0.235480$.

Once normalised to the total mass of the SSP, Eqn. \ref{eqn:ChabrierIMF} can be integrated over a certain mass range to tell us the number density ($n = N/V$) of stars in that mass range. Assuming that the IMF is the same everywhere, this is equivalent to the \textit{number} of stars in a $1 \textnormal{M}_{\textnormal{\astrosun}}$ SSP in a given mass range\footnote{Note that other authors, such as \citet{LPC02} and \citet{A10a}, choose to define the IMF as \textit{the mass} of stars in a $1 \textnormal{M}_{\textnormal{\astrosun}}$ SSP, $\Phi(M)$. This is related to the IMF defined in this work, $\phi(M)$, by $\Phi(M)=M\phi(M)$.}. This integrated, normalised IMF has units of $1/\textnormal{M}_{\textnormal{\astrosun}}$.

The Chabrier IMF predicts fewer stars of mass $< 1  \textnormal{M}_{\textnormal{\astrosun}}$ than the \citet{S55} IMF, and does so with a smoother transition than the multi-segment power-law \citet{K01} IMF. At masses above $1 \textnormal{M}_{\textnormal{\astrosun}}$, it has the same slope as the Kroupa IMF (an exponent of -2.3 in linear mass units, rather than the -2.35 used for the Salpeter IMF).

\subsection{Stellar lifetimes} \label{sec:Stellar lifetimes}
We adopt the metallicity-dependent lifetimes tabulated by \Citeauthor{P98} (1998, hereafter P98), kindly provided by R. Wiersma (priv. comm.). These account for stars in the mass range $0.6\leq M/\textnormal{M}_{\textnormal{\astrosun}} \leq 120$, and five different initial metallicities, from 0.0004 to 0.05 (where metallicity is the fraction $M_{\textnormal{Z}}/M$ here). The same study also provided SN-II yield tables, which we also use (see \S \ref{sec:SN-II yields}).

The lifetimes for different initial metallicities are plotted as a function of mass in Fig. \ref{fig:lifetimes}. Within the metallicity range shown, the most massive stars ($\sim 120 \textnormal{M}_{\textnormal{\astrosun}}$) live for up to $\sim3.3$ Myrs, depending on their initial metallicity, while the smallest stars that shed material during their lives ($\sim 0.85 \textnormal{M}_{\textnormal{\astrosun}}$) live for $\sim 10$ to 21 Gyrs. Stars of $\sim 1 \textnormal{M}_{\textnormal{\astrosun}}$ can live from $\sim 6$ to 10 Gyrs according to these lifetime tables, implying that some G V stars (also known as G dwarfs) would \textit{not} live for more than a Hubble time. The implications of this are briefly discussed in \S \ref{sec:Milky Way stars}.

\subsection{AGB wind yields} \label{sec:AGB wind yields}
We adopt the metallicity-dependent yield tables of \Citeauthor{M01} (2001, hereafter M01) for low- and intermediate mass stars, which eject their metals predominantly through stellar winds during their AGB phase.\footnote{Total yields from the RGB and AGB phases together are included in the M01 tables. For simplicity, we refer to these as `AGB wind' yields hereafter.} The SN-II yield tables of P98, which we also use, form a complete set with those of M01 for AGB winds. They are both based on the same Padova evolutionary tracks and do not require a large interpolation between them, as the AGB yields consider stars up to $5 \textnormal{M}_{\textnormal{\astrosun}}$ and the SN-II yields consider down to $7 \textnormal{M}_{\textnormal{\astrosun}}$.\footnote{We note that it is also possible to link the M01 AGB yields to the P98 SN-II yields at $6 \textnormal{M}_{\textnormal{\astrosun}}$, by including the P98 yields for electron-capture SNe (see P98,\S 4.2). Doing this makes a negligable difference to the results discussed in this work.} In this work, we consider the ejecta from AGB winds to occur at the end of a star's lifetime.

Fig. \ref{fig:agbwinds} shows the \textit{ejected mass} of metals (top left panel), total baryons (top right panel), and individual elements (bottom two panels) from AGB stars as a function of initial mass. This is different from the \textit{yield}, as it includes both the mass that passes through the stars unprocessed and any newly synthesised material.\footnote{The element `yield' of a star is defined as the mass of that element that is synthesised and ejected \citep{T80}. If an element undergoes a net destruction during stellar nucleosynthesis (e.g. hydrogen), then its yield will be negative, whereas the \textit{mass} of the element ejected will not.} The element abundances of the Sun from \citet{A09} are used to scale the amplitudes of the curves in Fig. \ref{fig:agbwinds}. 

We note that no elements heavier than oxygen present in the wind have been synthesised or destroyed in the AGB stars, but have instead been formed in previous generations of stars and pass through the AGB stars unprocessed. We have extrapolated the AGB wind yields from $5 \textnormal{M}_{\textnormal{\astrosun}}$ to $7 \textnormal{M}_{\textnormal{\astrosun}}$, so that they meet with the SN-II yields used. The exact position of this interface within the region $5 < M/\textnormal{M}_{\textnormal{\astrosun}} < 8$ does not significantly affect our results.

\begin{figure}
\centering
\includegraphics[totalheight=0.3\textheight, width=0.46\textwidth]{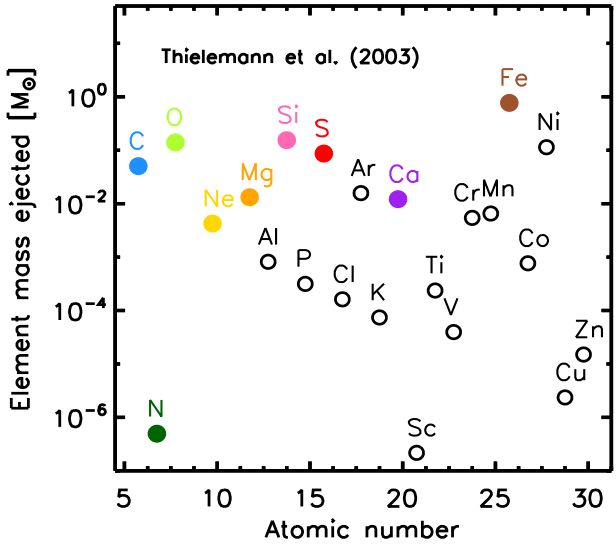}
\caption{The mass of each element ejected from SNe-Ia, according to the tabulation of \citet{T03}. Coloured circles represent elements that are considered in this work.}
\label{fig:SNIa}
\end{figure}

\subsection{SN-Ia yields} \label{sec:SN-Ia yields}
As with many other chemical enrichment models, we adopt the spherically symmetric `W7' model for our SN-Ia explosive yields, originally tabulated by \citet{N84}. We use a more recent iteration, by \Citeauthor{T03} (2003, hereafter T03). These tables provide the synthesised mass of forty two different element species. Unlike the AGB and SN-II yields, the SN-Ia yields used here are independent of the initial mass and metallicity of the progenitor system. The total mass ejected in a SN-Ia is assumed to be $1.23\textnormal{M}_{\textnormal{\astrosun}}$, the sum of the ejecta from the eleven elements considered in this work. As no H or He is ejected by SNe-Ia, this sum equals the mass of metals ejected. Fig. \ref{fig:SNIa} shows the ejected mass of each element. Iron is the most abundant, while there are also non-negligible amounts of oxygen, silicon and nickel.

SN-Ia yields that depend on the initial mass and metallicity of the progenitors are now also available in the literature (e.g. \citealt{S13}). We defer a study of the effect of such yields on our GCE model to future work.

Rather than make assumptions about the type and lifetimes of the progenitor systems involved, we instead use observationally-motivated DTDs to define the lifetimes of SN-Ia progenitors (see \S \ref{sec:SN-Ia DTD}).

\subsection{SN-II yields} \label{sec:SN-II yields}
Our preferred set of SN-II yields is tabulated by P98, and also kindly provided by R. Wiersma (priv. comm.). This set contains yields for initial masses ranging from 6 to 1000 $\textnormal{M}_{\textnormal{\astrosun}}$, and five initial metallicities from 0.0004 to 0.05. We only consider the existence of stars up to 120 $\textnormal{M}_{\textnormal{\astrosun}}$ here. Even then, the range provided by the P98 yields is significantly wider than, for example, the more commonly used yields of \citet{WW95}, which only go up to $40 \textnormal{M}_{\textnormal{\astrosun}}$.\footnote{Unlike the tables of \citet{WW95}, both sets of SN-II yield tables considered in this work account for the decay of nickel into iron shortly after the SN. P98 do so by simply adding the $^{56}$Ni yield to that of $^{56}$Fe, and \citet{CL04} by only tabulating yields $10^{8}$s after the explosion.} The P98 set also takes account of mass loss through winds prior to the SN. The inclusion of prior mass loss also effects the composition of the explosive yields, as we explain below.

\begin{figure*}
\centering
\includegraphics[totalheight=0.46\textheight, width=0.71\textwidth]{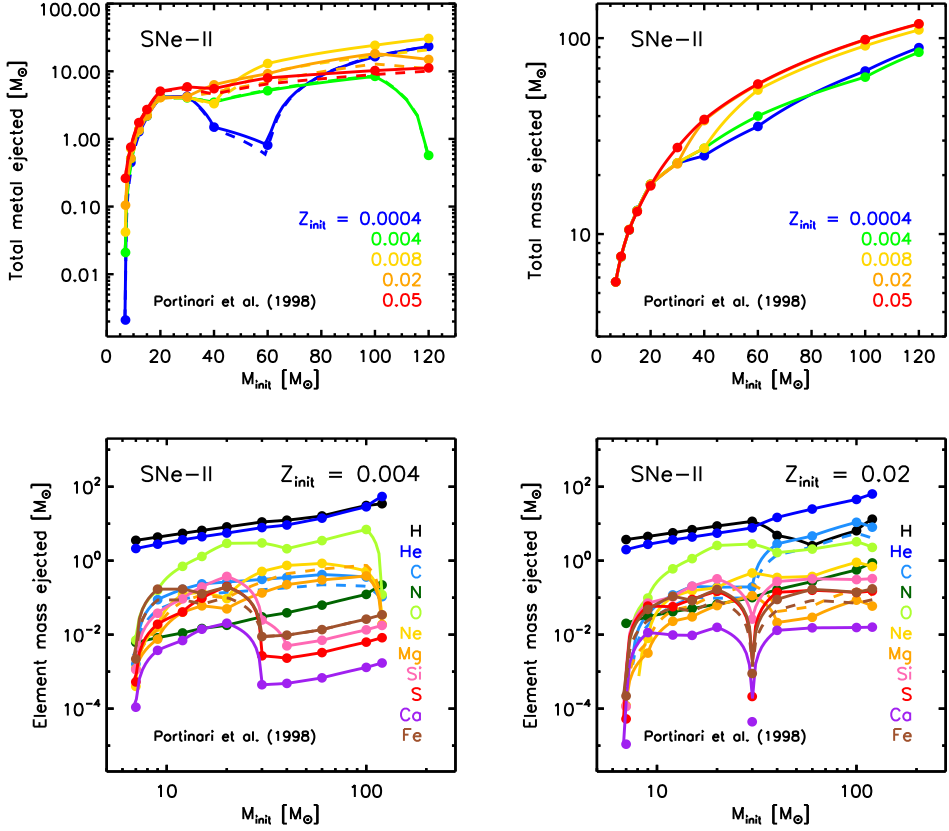}
\caption{Mass released by SNe-II from the \citet{P98} yield tables. Points indicate values from the yield tables. Solid lines indicate the interpolation used between these points. \textit{Top left}: The mass of metals ejected as a function of mass, for five different initial metallicities. \textit{Top right}: The total baryonic mass ejected as a function of mass, for five different initial metallicities. \textit{Bottom left}: The mass of each element ejected as a function of mass, for stars of $Z_{0}=0.004$. Dashed lines indicate the corrected C, Mg and Fe yields (see text). \textit{Bottom right}: Same as bottom left, for stars of $Z_{0}=0.02$.}
\label{fig:P98 SN-II}
\end{figure*}

Fig. \ref{fig:P98 SN-II} shows the \textit{ejected mass} of metals (top left panel), total baryons (top right panel), and individual elements (bottom two panels) from SNe-II as a function of initial mass, as is done for AGB winds in Fig. \ref{fig:agbwinds}. The dashed lines in Fig. \ref{fig:P98 SN-II} indicate corrections to the C, Mg and Fe yields that we include in our model, following the recommendation of \Citeauthor{W09b} (2009, see their \S A3.2). These \textit{ad hoc} corrections can be justified by uncertainties in the explosive yields tabulated by \citet{WW95}, on which the P98 SN-II yields are based. These corrections halve the yield of C and Fe and double the yield of Mg, relative to the originally tabulated values.

We note that the P98 yields show some sudden drops in the ejecta of certain elements. At low metallicities, the reduction in yield of the heaviest elements above $\sim 30 \textnormal{M}_{\textnormal{\astrosun}}$ is due to them being locked in the stellar remnant. Remnant masses increase significantly above $\sim 30 \textnormal{M}_{\textnormal{\astrosun}}$ at low metallicities due to low mass-loss efficiency prior to the SN. This effect is less severe for lighter elements, such as oxygen, as `pair creation' SNe are believed to dominate over `core collapse' SNe above $\sim 60 \textnormal{M}_{\textnormal{\astrosun}}$, allowing more of these elements to be ejected. At higher metallicities, more efficient mass loss prior to the SN inhibits large remnant formation. Increased mass loss at $Z_{0}\geq 0.02$ from massive, Wolf-Rayet stars also causes the larger He and C yields in this metallicity range. The removal of these elements in the wind in turn suppresses the explosive $\alpha$ element yields. For more details, see \S 5 of P98.

These specific features could have a signficant impact on our results. We therefore also test our GCE implementation with an alternative set of SN-II yields, that do not take account of prior mass loss, and therefore appear more stable as a function of initial mass and metallicity. This second set is taken from \Citeauthor{CL04} (2004, hereafter CL04). These account for stars of initial masses from 13 to 35 $\textnormal{M}_{\textnormal{\astrosun}}$, and so require both an extrapolation downwards to the upper mass limit for AGB winds (chosen here to be $7 \textnormal{M}_{\textnormal{\astrosun}}$), and upwards to a more reasonable maximum mass. We choose $M_{\textnormal{max}} = 120 \textnormal{M}_{\textnormal{\astrosun}}$ when using the CL04 SN-II yields in order to match the maximum mass considered for the P98 SN-II yields, and because such massive stars are known to exist and contribute to chemical enrichment in the real Universe. However, we caution that this represents a gross extrapolation into a regime well above that constrained by the original yield calculations. For this reason we use the CL04 yields only as a comparison to those of P98, in order to discern what effect prior mass loss might have on our overall results.

\section{The GCE Equation} \label{sec:The GCE Equation}
In this section, we present the GCE equations required to calculate the mass ejection rate from stars. The implementation of these equations into our semi-analytic model is described in \S \ref{sec:Implementation}.

Following the prescriptions given by \citet{T80}, the total rate of mass ejected by an SSP at time $t$ is given by

\begin{equation} \label{eqn:massejec}
e_{\textnormal{M}}(t) = \int^{M_{U}}_{M_{L}} (M-M_{\textnormal{r}}) \  \psi(t-\tau_{\textnormal{M}}) \  \phi(M) \  \textnormal{dM} \;\;,
\end{equation}
where $M$ is the initial mass of a star, $\tau_{\textnormal{M}}$ is its lifetime, $\psi(t-\tau_{\textnormal{M}})$ is the star formation rate when the star was born, $M_{\textnormal{r}}$ is the mass of the stellar remnant, and $\phi(M)$ is the normalised IMF by number, as given by Eqn. \ref{eqn:ChabrierIMF}.

$\psi(t-\tau_{\textnormal{M}}) \cdot \phi(M)$ gives us the birthrate of stars of mass $M$ at time $t-\tau_{\textnormal{M}}$. Multiplying this birthrate by $(M-M_{\textnormal{r}})$, the mass ejected by \textit{one} star of mass $M$, then gives us the total mass ejection rate by stars of mass $M$, at time $t$. We can then integrate this quantity over a suitable range of masses ($M_{L}$ to $M_{U}$) to obtain $e_{\textnormal{M}}(t)$.

The same equation can be written when only considering the \textit{metals} ejected by an SSP:

\begin{equation} \label{eqn:metalejec}
e_{\textnormal{Z}}(t) = \int^{M_{U}}_{M_{L}} M_{\textnormal{Z}}(M,Z_{0}) \  \psi(t-\tau_{\textnormal{M}}) \  \phi(M) \  \textnormal{dM} \;\;,
\end{equation}
where $M_{\textnormal{Z}} = y_{\textnormal{Z}}(M,Z_{0}) + Z_{0}\cdot(M-M_{\textnormal{r}})$ is the mass in metals returned to the gas phase by a star of mass $M$ (as clarified by \Citeauthor{M92} 1992, \S 4.1). This is made up of the mass- and metallicity-dependent yield\footnote{We define the metal yield as a mass $y_{\textnormal{Z}}$, rather than the mass fraction $p_{\textnormal{Z}}$ proposed by \citet{T80}, where $y_{\textnormal{Z}}=Mp_{\textnormal{Z}}$.} $y_{\textnormal{Z}}$, plus those metals present at the formation of the star that are later ejected unprocessed, $Z_{0}\cdot(M-M_{\textnormal{r}})$.

The same equation can be written again, when only considering individual chemical elements ejected by an SSP, replacing $M_{\textnormal{Z}}$ with $M_{\textnormal{i}}=y_{i}(M,Z_{0})+(M_{i}/M)(M-M_{r})$, the total mass of element $i$ returned to the gas phase by a star of mass $M$. However, for simplicity, we will proceed by describing the GCE equation in terms of the total metals ejected.

Eqn. \ref{eqn:metalejec} can be further split-up into four sub-components, representing the three modes of ejection, AGB winds, SNe-Ia and SNe-II:

\begin{align} \label{eqn:FinalGCEeqn}
\nonumber e_{\textnormal{Z}}(t)\ &\ = \ \int^{7 M_{\textnormal{\astrosun}}}_{0.85 M_{\textnormal{\astrosun}}} M_{\textnormal{Z}}^{\textnormal{AGB}}(M,Z_{0}) \  \psi(t-\tau_{\textnormal{M}}) \  \phi(M) \  \textnormal{dM} \\
\nonumber \ &\ + \ A^{'}\ k\ \int^{\tau_{0.85 M_{\textnormal{\astrosun}}}}_{\tau_{8 M_{\textnormal{\astrosun}}}}M_{\textnormal{Z}}^{\textnormal{Ia}}\ \psi(t-\tau)\ \textnormal{DTD}(\tau)\ \textnormal{d}\tau \\
\nonumber \ &\ +\  (1-A)\int^{16 M_{\textnormal{\astrosun}}}_{7 M_{\textnormal{\astrosun}}} M_{\textnormal{Z}}^{\textnormal{II}}(M,Z_{0}) \  \psi(t-\tau_{\textnormal{M}}) \  \phi(M) \  \textnormal{dM} \\
\ &\ +\  \int^{M_{\textnormal{max}}}_{16 M_{\textnormal{\astrosun}}} M_{\textnormal{Z}}^{\textnormal{II}}(M,Z_{0}) \  \psi(t-\tau_{\textnormal{M}}) \  \phi(M) \  \textnormal{dM} \;\;.
\end{align}
The first term in Eqn. \ref{eqn:FinalGCEeqn} represents the contribution to the ejected metals from AGB winds (approximating that the material is shed at the end of the stars' lives), with the symbols representing the same quantities as in Eqn. \ref{eqn:metalejec}. As can be seen, the integral extends to masses above the minimum mass of SN-Ia-producing binary systems ($\sim 3 M_{\textnormal{\astrosun}}$). Therefore, we are explicitly accounting for the ejection of metals during the AGB phase of such stars, prior to the SN.

The second term represents the contribution from SNe-Ia, parameterised with an analytic DTD motivated by observed SN-Ia rates (see \S \ref{sec:SN-Ia DTD}). Using a DTD means we do not have to make additional assumptions about the progenitor type of SNe-Ia, the binary mass function $\phi(M_{\textnormal{b}})$, secondary mass fraction distribution $f(\textnormal{M}_{2}/\textnormal{M}_{\textnormal{b}})$, or binary lifetimes in our modelling. These uncertain parameters become problematic when using the theoretical SN-Ia rate formalism of \citet{GR83}. The three SN-Ia DTDs that we consider in this work are described in \S \ref{sec:SN-Ia DTD}.

The coefficient $A^{'}$ in the second term of Eqn. \ref{eqn:FinalGCEeqn} gives the fraction of objects from \textit{the whole} IMF that are SN-Ia progenitors. This is subtly different from $A$ in the third term, which is the fraction of objects only in the mass range 3-$16 M_{\textnormal{\astrosun}}$ that are SN-Ia progenitors.\footnote{The use of the mass range 3 - $16 M_{\textnormal{\astrosun}}$ relates to the assumed mass range of SN-Ia-producing \textit{binary} systems in the single-degenerate scenario.} As clarified by \Citeauthor{A10a} (2010a, \S 3.3), these two coefficients are related by $A^{'}=A\cdot f_{3-16}$, where $f_{3-16}$ is the fraction of all objects in the IMF that have mass between 3 and $16 M_{\textnormal{\astrosun}}$. Our chosen value of $A$ is 0.028 (i.e. 2.8 per cent of the stellar systems in the mass range 3 - $16 M_{\textnormal{\astrosun}}$ are SN-Ia progenitors), as discussed in \S \ref{sec:Fiducial parameters}. The coefficient $k$ is given by 
\begin{equation}
k=\int^{M_{\textnormal{max}}}_{M_{\textnormal{min}}}\phi(M)\ \textnormal{d}M\;\;,
\end{equation}
and gives the number of stars in a $1 \textnormal{M}_{\textnormal{\astrosun}}$ SSP. For the Chabrier IMF used here, $f_{3-16}=0.0385$ and $k = 1.4772$ when assuming $M_{\textnormal{min}} = 0.1 \textnormal{M}_{\textnormal{\astrosun}}$ and $M_{\textnormal{max}} = 120 \textnormal{M}_{\textnormal{\astrosun}}$. 

The third term in Eqn. \ref{eqn:FinalGCEeqn} represents the ejection of metals, via SNe-II explosions, of all objects within the mass range $7.0 \leqslant M/\textnormal{M}_{\textnormal{\astrosun}} \leqslant 16.0$ that \textit{do not} produce SNe-Ia. Hence, the coefficient is $(1-A)$.\footnote{Note that, because the distribution of SN-Ia-producing binaries is assumed to follow the distribution of all objects, the value of $A$ is the same for any mass range within $3 < M/\textnormal{M}_{\textnormal{\astrosun}} < 16$.}

The fourth term represents the contribution to the ejection of metals from single, massive stars exploding as SNe-II.

We note here that Eqn. \ref{eqn:FinalGCEeqn} can also be rewritten so that all the modes of enrichment are expressed as time integrals, because the stellar lifetimes are a monotonic function of initial mass (e.g. P98, \S 8.7).

\begin{figure}
\centering
\includegraphics[totalheight=0.28\textheight, width=0.42\textwidth]{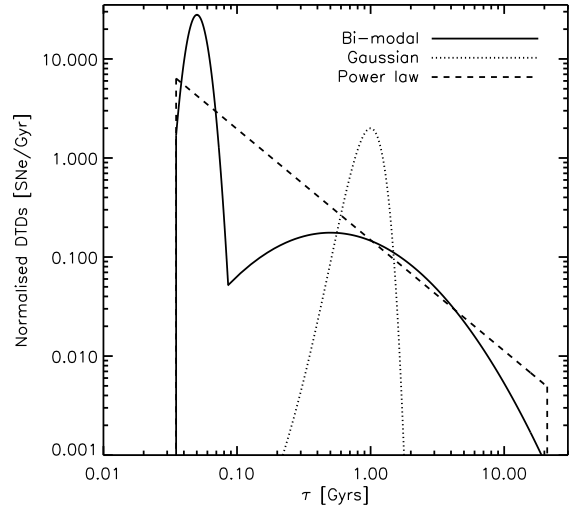}
\caption{The three SN-Ia delay-time distributions considered in this work. The dashed line corresponds to the power-law DTD given by Eqn. \ref{eqn:plDTD}. The dotted line corresponds to the narrow Gaussian DTD given by Eqn. \ref{eqn:ngDTD}. The solid line corresponds to the bi-modal DTD given by Eqn. \ref{eqn:ManDTD}. All three DTDs are normalised over the time range $\tau_{8 \textnormal{M}_{\textnormal{\astrosun}}}=35$ Myrs to $\tau_{0.85 \textnormal{M}_{\textnormal{\astrosun}}}=21$ Gyrs.}
\label{fig:DTDs}
\end{figure}

\subsection{SN-Ia delay-time distribution} \label{sec:SN-Ia DTD}
There have been many SN-Ia DTDs formulated in the literature. In this work, we consider three, shown in Fig. \ref{fig:DTDs}, and compare the results obtained from each.

The first is the power-law DTD with slope $-1.12$ proposed by \citet{MMB12}, formed from a fit to the SN-Ia rate derived from 66,000 galaxies (comprising 132 detected SNe-Ia) from the Sloan Digital Sky Survey II (SDSS-II):

\begin{equation} \label{eqn:plDTD}
\textnormal{DTD}_{\textnormal{PL}} = a(\tau/\textnormal{Gyr})^{-1.12}\;\;,
\end{equation}
where $\tau$ is the delay time since the birth of the SN-Ia-producing binary systems, and $a$ is the normalisation constant, taken here to be $a=0.15242\ \textnormal{Gyr}^{-1}$ (see Eqn. \ref{eqn:DTDnormalisation}). Similar power-law slopes have been suggested by a number of other works (e.g. \citealt{T08,MSG10,MM12}).

The second is the narrow, Gaussian DTD proposed by \citet{S04}, based on observations of 56 SNe-Ia in the range $0.2 < z < 1.8$ from the GOODS North and South fields. This form is given by

\begin{equation} \label{eqn:ngDTD}
\textnormal{DTD}_{\textnormal{NG}} = \frac{1}{\sqrt{2\pi\sigma_{\tau}^{2}}}\:e^{-\left( \tau-\tau_{c}\right) ^{2}/2\sigma_{\tau}^{2}}\;\;,
\end{equation}
where $\tau$ is again the delay time, $\tau_{c}=1$ Gyr is the characteristic time (on which the Gaussian distribution is centered), and $\sigma_{\tau}= 0.2\tau_{c}$ Gyrs is the characteristic width of the distribution.

The third is the bi-modal DTD proposed by \citet{MVP06}, motivated by simultaneously fitting both the observed SN-Ia rate and the distribution of SNe-Ia with galaxy B-K colour and radio flux, for a collection of samples over the redshift range $0.0 < z < 1.6$. This DTD includes a `prompt' component of SNe-Ia ($\sim54$ per cent of the total) that explode within $\sim$ 85 Myr of the birth of the binary, followed by a broader, delayed distribution. The \citet{MVP06} DTD has been expressed by \citet{M06} as

\begin{align} \label{eqn:ManDTD}
\nonumber \textnormal{log(DTD}_{\textnormal{BM}})& = \\
													\bigg{\{} & \begin{array}{ll}
													1.4-50(\textnormal{log}(\tau/\textnormal{yr})-7.7)^{2} & \textnormal{if } \tau < \tau_{0} \\
													-0.8-0.9(\textnormal{log}(\tau/\textnormal{yr})-8.7)^{2} & \textnormal{if } \tau > \tau_{0}
									 				\end{array} \;\;,
\end{align}
where $\tau$ is the delay time, and $\tau_{0}=0.0851$ Gyr is the characteristic lifetime separating the two components.

For all of these DTDs, the normalisation requirement is,

\begin{equation}\label{eqn:DTDnormalisation}
\int^{\tau_{\textnormal{max}}}_{\tau_{\textnormal{min}}}\textnormal{DTD}(\tau)\  \textnormal{d}\tau=1\;\;,
\end{equation}
where $\tau_{\textnormal{min}}=\tau_{8 \textnormal{M}_{\textnormal{\astrosun}}}$ and $\tau_{\textnormal{max}}=\tau_{0.85 \textnormal{M}_{\textnormal{\astrosun}}}$ are the minimum and maximum assumed lifetimes of a SN-Ia-producing binary in the single-degenerate scenario (i.e. the lifetimes of the largest and smallest possible secondary stars), respectively. Strictly, their values depend on the stellar lifetime tables used, and therefore also on the metallicity of the stars. However, we choose to fix their values for Eqn. \ref{eqn:DTDnormalisation} to those provided by the P98 lifetime tables for stars of $Z_{0}=0.02$, namely $\tau_{\textnormal{min}}=35$ Myrs and $\tau_{\textnormal{max}}=21$ Gyrs. Our chosen value of $\tau_{\textnormal{min}}=35$ Myrs is in line with those commonly used in the literature, with assumed values ranging from $\sim 30$ Myrs (e.g. \citealt{MG86,PM93,MR01,M09}) to $\sim 40$ Myrs (e.g. \citealt{Gr05}). This choice also means that $\sim 48$ per cent of SNe-Ia explode within 400 Myrs when using the power-law DTD, which is close to the $\sim 50$ per cent predicted from observations of the SN-Ia rate by \citet{B10}, and around the lower limit determined from SN remnants in the Small and Large Magellanic Clouds by \citet{MB10}.

\section{Implementation} \label{sec:Implementation}
The GCE equation given by Eqn. \ref{eqn:FinalGCEeqn} has been implemented into our semi-analytic model so that the mass of chemical elements ejected is calculated at each simulation timestep. The key aspects of this implementation are outlined in the following sub-sections.
 
\subsection{SFH, ZH and EH arrays} \label{sec:SFH, ZH and EH arrays}
There are three galaxy-dependent values that are required for us to predict ejection rates from stars: the star formation history (SFH), the total gas-phase metallicity history (ZH) and the gas-phase element abundance history (EH). The SFH is required to identify $\psi(t-\tau_{\textnormal{M}})$, the ZH is required to identify $Z_{0}$, and the EH is required to calculate the unprocessed ejecta of each individual element within the semi-analytic model.\footnote{Although the total gas-phase metallicity history could be derived by simply summing the element abundances, keeping two separate history arrays gives us the freedom to vary the number of chemical elements we choose to track, and also to easily record the relative contribution of the three ejection modes to the total metal production.} We accommodate these histories into arrays in our code.

\begin{figure}
\centering
\includegraphics[totalheight=0.25\textheight, width=0.35\textwidth]{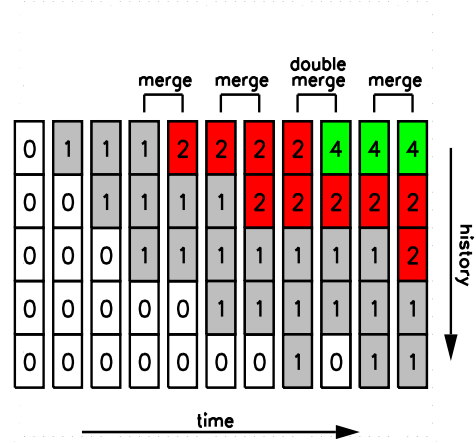}
\caption{The evolution of the first five bins (rows) of a history array for an isolated galaxy. The numbers represent the time width of a bin in units of one timestep. At every timestep in the code (columns, moving left to right), a new bin is `activated'. Active bins are coloured in the schematic (grey for single-width bins, red for double-width bins, and green for quadruple-width bins). When three or more active bins have the same width, two of the bins are immediately merged, as indicated at the top of the schematic.}
\label{fig:SFHbins}
\end{figure}

The \textsc{L-Galaxies} time structure is made up of 63 snapshots (when run on the \textsc{Millennium} simulation), each containing 20 timesteps. As there are nearly 26 million galaxies by $z=0$ in the semi-analytic model, it would require a significant amount of memory for us to store the full histories of each galaxy at the resolution of one timestep. Therefore, we instead take a more dynamic approach; each galaxy has SFH, ZH and EH arrays of only 20 array-elements (hereafter, time `bins'). As time elapses in the simulation, the width of older bins (those storing data from higher redshifts) increases, while new bins are `activated' with a default width of one timestep. Thus, the whole history of each galaxy can be stored with a time resolution that decreases with lookback time. High precision at recent times is especially important when calculating galaxy luminosities as a post-processing step, as young stars from recent star formation episodes tend to dominate the light. The evolution of these history arrays with time is illustrated in the schematic in Fig. \ref{fig:SFHbins}. We have checked that changing the number of bins in the history arrays does not affect the chemical evolution in the model by testing our model with a range of history bin resolutions including full resolution (i.e. $63\times20$ bins per galaxy).

\begin{figure}
\centering
\includegraphics[totalheight=0.45\textheight, width=0.47\textwidth]{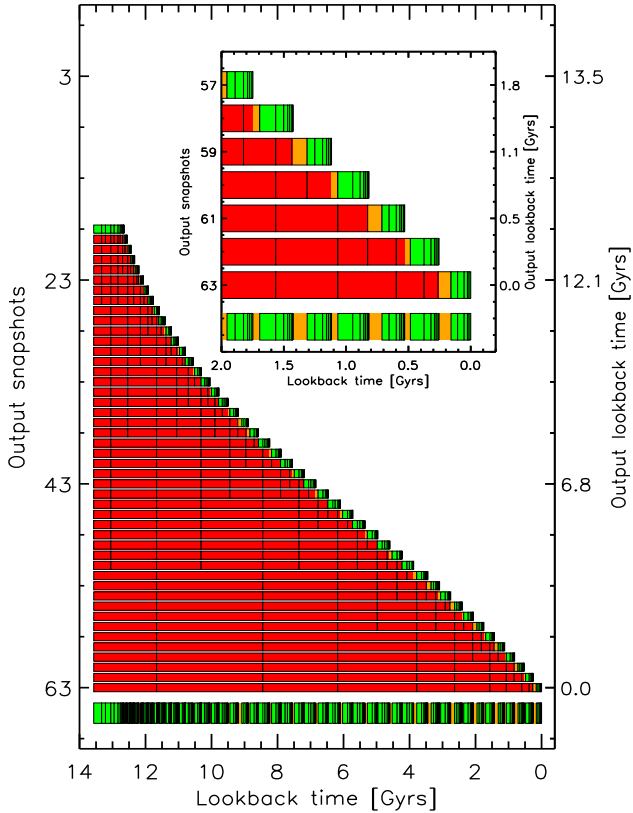}
\caption{Schematic illustrating how history arrays are `stitched together' in post processing to form higher-resolution histories when plotting data. At every output snapshot (y axis), a galaxy has a series of history bins (black boxes). The most recent bins from each output (in green) are extracted and used to form a higher-resolution, non-overlapping history (shown in the bottom row). The other bins (in red) are discarded. So that there are no gaps in the reconstructed histories, a fraction of the mass from partially overlapping bins is also included (in orange). This means that many hundreds of bins (depending on the formation time of the galaxy) can be used to make plots, rather than only 20 from the $z=0$ history. The inlay shows a zoom-in of the bottom-right region of the main schematic.}
\label{fig:stitched_SFHs}
\end{figure}

By $z=0$, the older bins in such histories can be up to $\sim 3$ Gyrs wide. This is acceptable when calculating the chemical enrichment within the code, as the bins are integrated over more finely at each timestep (see \S \ref{sec:Implementing the GCE equation}). However, when \textit{plotting} relations using only the output $z=0$ history bins, the lower resolution at high-$z$ does not correctly represent the smooth chemical evolution actually occuring in our model. In these cases (for example, the [Fe/H]-[O/Fe] relation in Fig. \ref{fig:FeH-OFe_SingleMW_AllDTDs}), we construct higher resolution histories as a post-processing step, by `stitching together' the highest-resolution bins from the histories of \textit{all} output snapshots, rather than just those from $z=0$. This procedure is illustrated by the schematic in Fig. \ref{fig:stitched_SFHs}. In this way, a much smoother evolution can be plotted, which more accurately represents the chemical evolution occuring within the code. We note that when doing this for the disc components of galaxies, account needs to be taken of stars that move from the disc to the bulge through disc instabilities, by ensuring that the total mass formed in the stitched-together bins does not exceed the mass formed in the $z=0$ history bins over the same time span.

\subsection{Implementing the GCE equation}\label{sec:Implementing the GCE equation}
In order to model GCE, Eqn. \ref{eqn:metalejec} needs to be implemented into \textsc{L-Galaxies} as an algorithm, involving numerical integration and interpolation between values in a number of look-up tables. All non-model-dependent terms (i.e. everything except the SFHs, ZHs and EHs) are pre-calculated and stored in look-up tables, in order to speed-up the runtime of the code. This is possible because the time structure of the history arrays is, by construction, the same for all galaxies at any given time. Therefore, we can know \textit{a priori} the range of masses of stars that will explode in any given timestep.

We can re-write Eqn. \ref{eqn:metalejec} as

\begin{equation} \label{eqn:metalejecalgorithm1}
e_{\textnormal{Z}}(t) = \psi(t-\tau) \left[ \int^{M_{U}}_{M_{L}} M_{\textnormal{Z}}(M,Z_{0})\cdot\phi(M)\  \textnormal{dM} \right]\;\;, 
\end{equation}
where $\psi(t-\tau)$ can be put outside the integral if we assume, for each thin strip of the SFH integrated, that all the stars of mass $M_{L} \leq M \leq M_{U}$ are born \textit{at the same time} (i.e. $\tau_{M_{L}}=\tau_{M_{U}}=\tau$). We can then pre-calculate the integral in Eqn. \ref{eqn:metalejecalgorithm1} numerically to obtain a value for each initial metallicity, in every history bin of every timestep that the semi-analytic model will run through, and store it in a 3-dimensional look-up table. The true value of $Z_{0}$ for a given galaxy is then used within the semi-analytic model to interpolate between these pre-calculated results at each timestep, and $\psi(t-\tau)$ is multiplied-in. The total mass in metals ejected is then given by $e_{\textnormal{Z}}(t)\cdot \Delta t$, where $\Delta t$ is the width of the timestep. The same procedure is used to obtain the total mass ejected, and the total amount of each chemical element ejected at each timestep. Once the ejected masses are calculated, we transfer the material to either the galaxy's ISM or CGM, as described in \S \ref{sec:SN feedback}.

\subsection{Infall, Cooling and Outflows}\label{sec:Infall, Cooling and Outflows}
The chemical enrichment recipe outlined above is only part of the relevant physics needed to accurately model the chemical evolution of galaxies. The distribution of these metals between the various components of galaxies, and out into the IGM, as well as the infall and cooling of gas, are also important considerations when looking beyond a simple closed-box model. Treatments of these physical processes are already incorporated into \textsc{L-Galaxies}, as described by \citet{G10}. A brief outline is also given below.

We note here that \textsc{L-Galaxies} considers three classes of galaxy: those at the centre of a main DM halo, also known as a `friends-of-friends (FOF) group' (type 0 galaxies), those at the centre of their own DM subhalo but not of their associated FOF group (type 1 galaxies), and those galaxies that have lost their DM subhalo through tidal disruption but have not yet merged with a central galaxy or been tidally disrupted themselves (type 2 galaxies). The prescriptions for the physical processes included in the model are then applied to galaxies according to their type. For example, infall of pristine gas is only allowed to occur for type 0 galaxies, whereas stripping of hot gas can only occur for type 1 galaxies (once they are within the virial radius of the central galaxy).

\subsubsection{Infall}
The mass of pristine gas (assumed to be 75 per cent hydrogen and 25 per cent helium) infalling onto the DM halo is simply determined by the difference between the assumed baryon fraction $f_{b}$ and the actual baryon fraction $M_{b}/M_{\textit{\textnormal{DM}}}$ in the DM halo. The assumed baryon fraction is reduced from the cosmic baryon fraction $f_{b,\textnormal{cos}}$ (assumed to be 0.17, as given by WMAP1) due to reionisation, and is parameterised following \citet{G00} as

\begin{equation}
f_{b}(z,M_{\textnormal{vir}}) = f_{b,\textnormal{cos}}\bigg{[}1+(2^{2/3}-1)\bigg{(}\frac{M_{\textnormal{vir}}}{M_{\textnormal{c}}(z)}\bigg{)}^{-2}\bigg{]}^{-3/2}\;\;,
\end{equation}
where $M_{\textnormal{vir}}$ is the virial mass of the DM halo, and $M_{\textnormal{c}}(z)$ is the chosen charateristic halo mass, whose dependence on redshift has been calculated by \citet{OGT08}. In this formalism, $f_{b}$ tends towards $f_{b,\textnormal{cos}}$ as $M_{\textnormal{vir}}$ increases. Pre-enriched gas can also be re-accreted onto the DM haloes of central galaxies, in addition to this pristine infall (see \S \ref{sec:SN feedback}).

\subsubsection{Cooling}
Following \citet{WF91}, the cooling of gas from the CGM onto the disc is considered to fall into two regimes; at early times and in low-mass DM haloes, gas is able to cool rapidly in less than the free-fall time, with the cold-flow accretion onto the central galaxy modelled as

\begin{equation}
\dot{M}_{\textnormal{cool}} = \frac{M_{\textnormal{acc}}}{t_{\textnormal{dyn,h}}}\;\;,
\end{equation}
where $M_{\textnormal{acc}}$ is the mass of gas accreted onto the DM halo, and the dynamical time of the DM halo is $t_{\textnormal{dyn,h}} = R_{\textnormal{vir}}/V_{\textnormal{vir}}=0.1H(z)^{-1}$. At late times and in massive DM haloes, the accretion shock radius is large, leading to the formation of a hot gas atmosphere. In this case, the accretion rate onto the central galaxy is reduced to

\begin{equation}
\dot{M}_{\textnormal{cool}} = \frac{r_{\textnormal{cool}}}{R_{\textnormal{vir}}}\frac{M_{\textnormal{hot}}}{t_{\textnormal{dyn,h}}}\;\;.
\end{equation}
Here, the cooling radius $r_{\textnormal{cool}}$ is set by the cooling function of \citet{SD93}, and $M_{\textnormal{hot}}$ is the mass of shocked gas in the hot gas reservoir (see \citealt{G10}, \S 3.2). This accreted gas is then able to form stars, following a simplified form of the Kennicutt-Schmidt law (see \citealt{G10}, \S 3.4).

\subsubsection{SN feedback} \label{sec:SN feedback}
SNe explosions can reheat cold gas and also eject it from the DM halo of galaxies. In previous versions of \textsc{L-Galaxies}, the amount of energy released by SNe was assumed to be proportional to the mass of stars \textit{formed} $\Delta M_{*}$ at that time. Now that we have discarded the instantaneous recycling approximation, it is more appropriate to relate this energy to the mass of material \textit{released} by stars at that time. The total amount of energy produced by SN feedback is therefore parameterised as

\begin{equation}
E_{\textnormal{SN}} = \epsilon_{\textnormal{h}}\cdot{}\frac{1}{2} e_{\textnormal{M}}(t)\Delta t V^{2}_{\textnormal{SN}}\;\;,
\end{equation}
where $\epsilon_{\textnormal{h}}$ is the halo-velocity-dependent SN energy efficiency, $e_{\textnormal{M}}(t)\cdot \Delta t$ is the mass released by stars in one timestep (see Eqn. \ref{eqn:massejec}), and $V_{\textnormal{SN}}$ is the SN ejecta speed, assumed to be fixed at 630 km/s. This differs from the prescription used by \citet{G10} in the use of $e_{\textnormal{M}}(t)\cdot \Delta t$ rather than $\Delta M_{*}$. Due to this change, we have doubled the value of $\epsilon_{\textnormal{h}}$, in order to have the same total SN feedback energy ($E_{\textnormal{SN}}$) as previously used in the model. A thorough investigation into the precise values of model parameters required following our new GCE implementation is reserved for future work.

In our new, default GCE implementation, all stars dying in the stellar disc release material and energy into the ISM, whereas stars dying in the bulge and stellar halo release material and energy into the hot CGM. The energy dumped into the ISM by disc stars can then be used to reheat and possibly eject some (fully mixed) cold gas. Energy dumped into the CGM can also contribute to ejection. The amount of gas ejected from the DM halo into an external reservoir is given by

\begin{equation}
\Delta M_{\textnormal{ejec}}=\frac{E_{\textnormal{SN}}-\frac{1}{2}\epsilon_{\textnormal{disc}} e_{\textnormal{M}}(t)\Delta t V^{2}_{\textnormal{vir}}}{\frac{1}{2}V^{2}_{\textnormal{vir}}}\;\;,
\end{equation}
where $\epsilon_{\textnormal{disc}}\cdot e_{\textnormal{M}}(t)\cdot \Delta t $ is the amount of gas that is reheated but does not escape the potential well. The ejected gas is then allowed to return to the DM halo over timescales that are proportional to $V_{\textnormal{vir}}/t_{\textnormal{dyn,h}}$.\footnote{\citet{H13} have found that scaling the reincorporation time to the inverse of the DM halo mass allows the semi-analytic model to better reproduce the evolution of the galaxy stellar mass and luminosity functions with redshift. We will incorporate this improvement with our new GCE model in future work.} This constitutes a second component of gas infall that has been pre-enriched by the galaxy.

We have also implemented an alternative feedback prescription which includes metal-rich winds. These winds dump some material released by disc SNe-II directly into the hot gas. This scheme is discussed in \S \ref{sec:Galactic winds}.

\subsection{Default set-ups} \label{sec:Fiducial parameters}
There can be many free parameters involved when developing a chemical enrichment model. We have limited ourselves to only one new free parameter: the fraction, $A$, of objects in an SSP in the range $3\leq M/\textnormal{M}_{\textnormal{\astrosun}}\leq 16$ that are SN-Ia progenitors.\footnote{We note again that the SN efficiency parameter $\epsilon_{\textnormal{h}}$ has also been modified to ensure that the total SN feedback energy is unchanged (see \S \ref{sec:SN feedback}). All other model parameters have been kept to the values used by \citet{G10}.} $A$ is specifically `tuned' so that the peak of the [Fe/H] distribution for G dwarfs in our MW-type galaxy sample is around the solar value (see \S \ref{sec:Milky Way stars}). An increase in $A$ corresponds to an increase in [Fe/H], and we find that the best value is $A\sim0.028$ for all three of the DTDs we consider (see \S \ref{sec:SN-Ia DTD}). A single value of $A$ was also found to be suitable for a range of different SN-Ia DTDs by \citet{M09}.  All other results discussed in this work are obtained without further tuning. In the following, we label results using the bi-modal, power-law, and narrow Gaussian DTDs with `BM', `PL' and `NG', respectively.

We note that our preferred value of $A = 0.028$ is similar to that commonly found in the literature. For example, \citet{Gr05} took a value of $A^{'}=0.001$ when also using a Chabrier IMF, which equates to a value of $A=0.026$.  Similarly, \citet{dP07} take a preferred value of 0.027 for a Kroupa IMF (assuming a SN-Ia progenitor mass range of $1.5-10 \textnormal{M}_{\textnormal{\astrosun}}$). \citet{A10a} allow for a value between 0.015 and 0.05, preferring 0.03 when using a slightly top-heavy Chabrier IMF (i.e. a slope of 2.15 rather than 2.3 for $M > 1 \textnormal{M}_{\textnormal{\astrosun}}$, see Eqn. \ref{eqn:ChabrierIMF}). Other works, which have used IMFs with a smaller fraction of stars above $1 \textnormal{M}_{\textnormal{\astrosun}}$, have taken slightly higher values. For example, \citet{MR01} and \citet{F04} prefer $A=0.05$ when using a \citet{S86} IMF. \citet{CM09} and \citet{M06} take values of $A^{'}=0.0020$ and $0.0025$ for a Scalo IMF, which corresponds to $A\sim0.05$ and $0.06$, respectively. And P98 find $A=0.05$ - $0.08$ when using a Salpeter IMF (and a SN-Ia progenitor mass range of $3-12 \textnormal{M}_{\textnormal{\astrosun}}$).

\begin{figure}
\centering
\includegraphics[totalheight=0.32\textheight, width=0.46\textwidth]{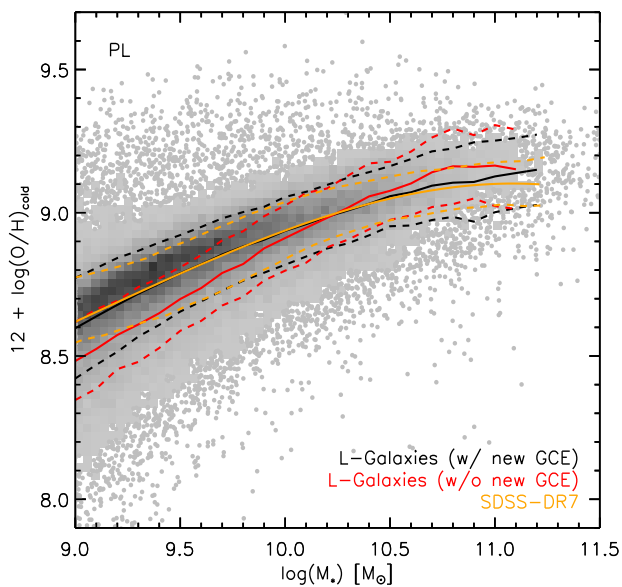}
\caption{The $M_{*}$-$Z_{\textnormal{cold}}$ relation (where $Z_{\textnormal{cold}}=12+\textnormal{log}(N_{\textnormal{O}}/N_{\textnormal{H}})$) for \textsc{L-Galaxies} with the new GCE implementation and using a power-law SN-Ia DTD (points and black lines). This relation is compared to that of \textsc{L-Galaxies} prior to the new GCE implementation (red lines), and a fit to the observed $M_{*}$-$Z_{\textnormal{g}}$ relation for emission-line galaxies from the SDSS-DR7 (orange lines) by \citet{YKG12}.}
\label{fig:GasMZR}
\end{figure}

Our chosen value of $A\sim 0.028$ is also in line with expectations from observations of the SN-Ia rate, with the fraction of SN-Ia-producing stars in the range 3 - $8 \textnormal{M}_{\textnormal{\astrosun}}$ believed to be between 0.03 and 0.1 \citep{MM12}. For the range 3 - $16 \textnormal{M}_{\textnormal{\astrosun}}$, this equates to between $\sim 0.024$ and 0.081 (for a Chabrier IMF).

\section{Results} \label{sec:General Results}
In the following sub-sections, we compare results from our updated semi-analytic model to observational data for local star-forming galaxies (\S \ref{sec:The mass-metallicity relation}), Milky Way disc stars (\S \ref{sec:Milky Way stars}) and local elliptical galaxies (\S \ref{sec:Ellipticals}). In doing so, we are attempting both to assess the success of our GCE implementation and to further constrain which of the SN-II yield tables and SN-Ia DTDs described in \S \ref{sec:GCE ingredients} and \S \ref{sec:SN-Ia DTD} perform best across the range of data considered. In what follows, `element enhancement' refers to the ratio of element \textit{x} to iron, [\textit{x}/Fe], and `element abundance' refers to the ratio of element \textit{x} to hydrogen, [\textit{x}/H].\footnote{The element ratios discussed in this work are normalised to solar values, using the following equation: $[x/y] = \textnormal{log}(M_{x}/M_{y}) - \textnormal{log}(M_{x\textnormal{\astrosun}}/M_{y\textnormal{\astrosun}})$. Note that $M_{x}/M_{y} = (A_{x}/A_{y})\cdot (\epsilon_{x}/\epsilon_{y})$, where $A_{x}$ is the atomic weight of element \textit{x}, $\textnormal{log}(\epsilon_{x})=\textnormal{log}(n_{x}/n_{\textnormal{H}})+12$ is the abundance of element \textit{x}, and $n_{x}$ is the number density of atoms of element \textit{x}. For hydrogen, $A_{\textnormal{H}}=1.008$ and $\textnormal{log}(\epsilon_{\textnormal{H}})=12.0$.} Throughout this work, we normalise our model values to the set of solar abundances used for the observations to which we compare. For clarity, we have selected a representative sample of $\sim 480000$ $z=0$ galaxies and their progenitors for the plots in this section.

\subsection{The mass-metallicity relations} \label{sec:The mass-metallicity relation}
One of the key diagnostics used to analyse the chemical evolution of galaxies is the relation between their stellar mass ($M_{*}$) and gas-phase metallicity ($Z_{\textnormal{g}}$). The large statistical power of the SDSS allowed \citet{T04} to determine the $M_{*}$-$Z_{\textnormal{g}}$ relation for emission-line galaxies in the local Universe. They found a clear positive correlation below $\sim 10^{10.5} \textnormal{M}_{\textnormal{\astrosun}}$ with a $1\sigma$ scatter of only 0.1 dex. Above this mass, the relation was found to flatten. Here, we compare our $z=0$ model mass-metallicity relations for gas and stars with those observed. We also have the opportunity to directly compare \textsc{L-Galaxies} results before and after the new GCE implementation -- something we are not able to do when discussing individual element ratios in later sub-sections.

\begin{figure}
\centering
\includegraphics[totalheight=0.32\textheight, width=0.46\textwidth]{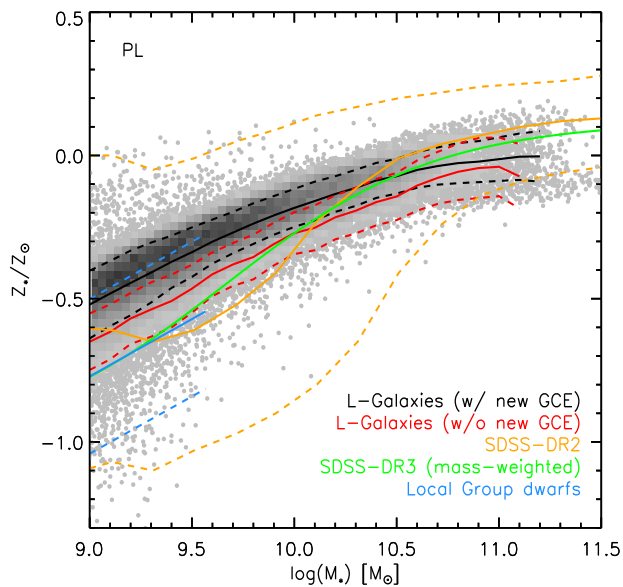}
\caption{The $M_{*}$-$Z_{*}$ relation (where $Z_{*}=\textnormal{log}(M_{*,\textnormal{Z}}/M_{*}/0.02)$) for \textsc{L-Galaxies} with the new GCE implementation and using a power-law SN-Ia DTD (points and black lines). This relation is compared to that of \textsc{L-Galaxies} prior to the new GCE implementation (red lines), the observed relation from the SDSS-DR2 (orange lines) by \citet{G05}, a fit to the mass-weighted relation from the SDSS-DR3 (green line) by \citet{P08}, and to a set of Local Group dwarf galaxies (blue lines) by \citet{WCD08}.}
\label{fig:StellarMZR}
\end{figure}

Fig. \ref{fig:GasMZR} shows the $M_{*}$-$Z_{\textnormal{cold}}$ relation for \textsc{L-Galaxies} with the new GCE implementation and using the power-law DTD (points and black lines). 95600 model galaxies were selected such that $\textnormal{log}(M_{*}) \geq 8.6$ and $-2.0 \leq \textnormal{log(SFR)} \leq 1.6$, in order to match the dynamic range of the SDSS-DR7 observations. We note here that both the gas-phase and stellar mass-metallicity relations are very similar for all three of the DTDs considered. This is because the SN-Ia DTD has little impact on the abundance of oxygen, which is the most abundant heavy element and is produced predominantly by SNe-II.

In Fig. \ref{fig:GasMZR} we also plot a fit to the same relation for \textsc{L-Galaxies} \textit{prior} to the new GCE implementation (red lines), and a fit to the observed $M_{*}$-$Z_{\textnormal{g}}$ relation from the SDSS data release 7 (SDSS-DR7) (orange lines).\footnote{This fit to the SDSS-DR7 is given by $26.6864 - 6.63995\;\textnormal{log}(M_{*}) + 0.768653\;\textnormal{log}(M_{*})^{2} - 6.0282147\;\textnormal{log}(M_{*})^{3}$, and is an updated version of the SDSS-DR2 relation from \citet{T04}, using twice as many galaxies \citep{YKG12}.} We can see that there is very good agreement between the observations and our new model at $z=0$. Both the slope and amplitude of the new model relation are in better agreement with observations than those of the previous model. The increase in amplitude at lower mass is due to a) our new GCE implementation (i.e. the input yields) allowing a different amount of metal into the ISM than the fixed 3 per cent yield assumed before, and b) our new SN feedback scheme allowing more oxygen to stay in the ISM after it is released by stars, rather than being instantly `reheated' into the CGM. This is because the energy input by a population of SNe is now distributed over time, rather than all dumped at once into the ISM straight after star formation, when a lot of oxygen is also released (see \S \ref{sec:SN feedback}).

The scatter of our new model $M_{*}$-$Z_{\textnormal{g}}$ relation is slightly larger than that seen in the SDSS. Studying the properties of outliers above and below the $M_{*}$-$Z_{\textnormal{g}}$ relation can tell us a lot about the evolution of galaxies (e.g. \citealt{DGK07,PPS08,Z12a}). We defer a detailed analysis of such galaxies in our model to later work.

We note here that the gas-phase metallicity is now defined as $Z_{\textnormal{cold}}=12+\textnormal{log}(N_{\textnormal{O}}/N_{\textnormal{H}})$ in our new model, in exactly the same way as in observations, where $N_{\textnormal{O}}$ and $N_{\textnormal{H}}$ are the number of atoms of oxygen and hydrogen, respectively. Previously, the approximation $Z_{\textnormal{cold}}=9.0+\textnormal{log}(M_{\textnormal{Z,cold}}/M_{\textnormal{cold}}/0.02)$ was used, where 9.0 was the assumed solar oxygen abundance and 0.02 the assumed solar metallicity. The difference in the value obtained when using these two methods is only small, with the new formulation estimating a metallicity $\sim 0.04$ dex lower than the old formulation.

Fig. \ref{fig:StellarMZR} shows the $z=0$ relation between the stellar mass and stellar metallicity ($Z_{*}$) of our model galaxies (using a power-law DTD), after our new GCE implementation (points and black lines), and prior to it (red lines). In both cases, solar-normalised metallicities are calculated as $Z_{*}=\textnormal{log}(M_{*,\textnormal{Z}}/M_{*}/0.02)$, using the same solar metallicity of $Z_{\textnormal{\astrosun}}=0.02$ assumed in the stellar population synthesis models that obtained stellar metallicities in the SDSS-DR2 (A. Gallazzi, priv. comm.).

Below $M_{*} = 10^{10.5}\textnormal{M}_{\textnormal{\astrosun}}$, the new model $M_{*}$-$Z_{*}$ relation is similar in shape to that of the previous model, but with an amplitude $\sim 0.1$ dex higher. This is also higher than observed at low mass (although this is a region where observations are are not well constrained). The mass-weighted $M_{*}$-$Z_{*}$ relation of \citet{P08} (green line) from the SDSS-DR3 probably provides the best comparison with our model, as we also consider mass-weighted metallicities. The \citet{P08} relation also shows good correspondance with observations of Local Group dwarfs by \citet{WCD08} (blue lines). We can see that the general trend of decreasing $Z_{*}$ with $M_{*}$ is reproduced in our model, despite low-$M_{*}$, star-forming model galaxies being too metal-rich by $z=0$.

\citet{HT10} have shown that a more realistic treatment of stellar disruption, whereby satellite galaxies have their stellar component \textit{gradually} stripped, can help steepen the slope of the $M_{*}$-$Z_{*}$ relation in semi-analytic models. This could bring the low-mass end of our model relation into better agreement with observations. Including such a gradual disruption scheme into \textsc{L-Galaxies} will be the focus of future work.

The model $M_{*}$-$Z_{\textnormal{cold}}$ and $M_{*}$-$Z_{*}$ relations when using the CL04 SN-II yields have slightly shallower slopes and are $\sim 0.1$ dex higher than those assuming the P98 yields. They therefore have a higher amplitude than observed. This is because the CL04 yield set allows more oxygen to be produced and ejected from stars when extrapolated to $120 \textnormal{M}_{\textnormal{\astrosun}}$, particularly at low metallicity.

\textit{} \newline
To conclude this section, we can say that our new GCE implementation improves the correspondance between our model and observations of gas-phase metallicities in local, star-forming galaxies. This was by no means a foregone conclusion, considering the significant changes to the chemical evolution modelling we have implemented. However, further improvement to the semi-analytic model is still required in order to better match the observed total \textit{stellar} metallicities of galaxies at $z=0$.

\begin{figure}
\centering
\includegraphics[totalheight=0.2\textheight, width=0.46\textwidth]{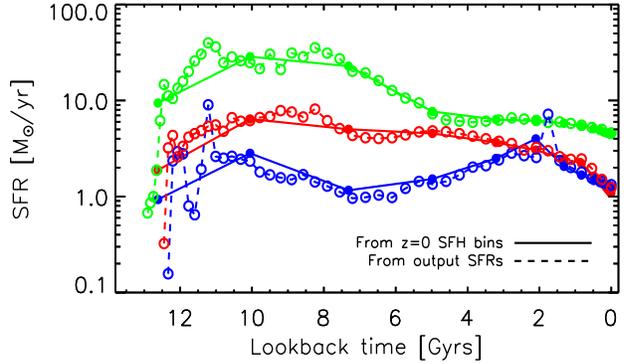}
\caption{Three example SFHs from our MW-type model sample. Filled circles represent the histories as recorded by the 20 SFH bins at $z=0$. Open circles represent the SFRs at every output snapshot of the simulation.}
\label{fig:MW_SFHs}
\end{figure}

\begin{figure}
\centering
\includegraphics[totalheight=0.26\textheight, width=0.38\textwidth]{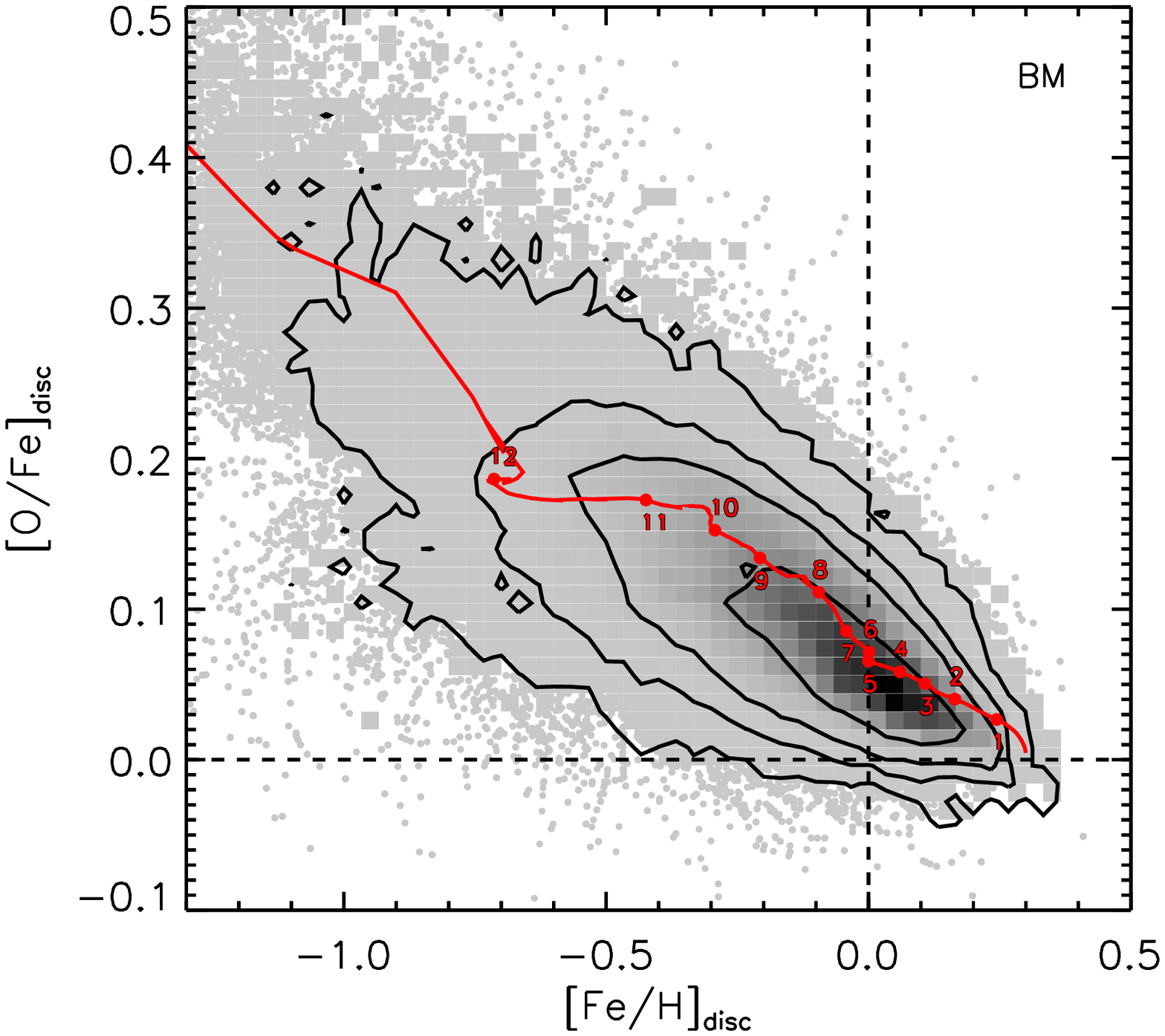} \\
\includegraphics[totalheight=0.26\textheight, width=0.38\textwidth]{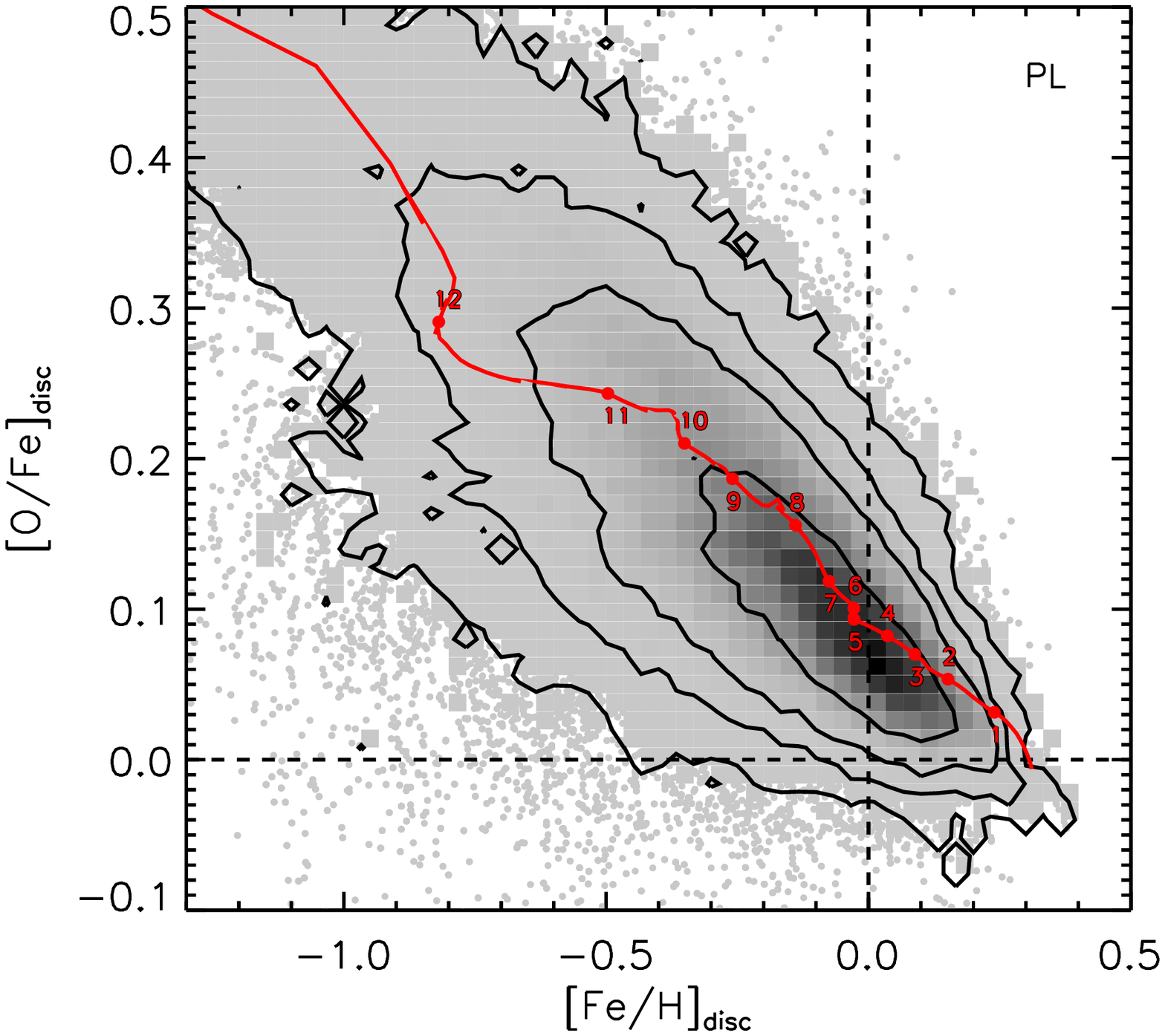} \\
\includegraphics[totalheight=0.26\textheight, width=0.38\textwidth]{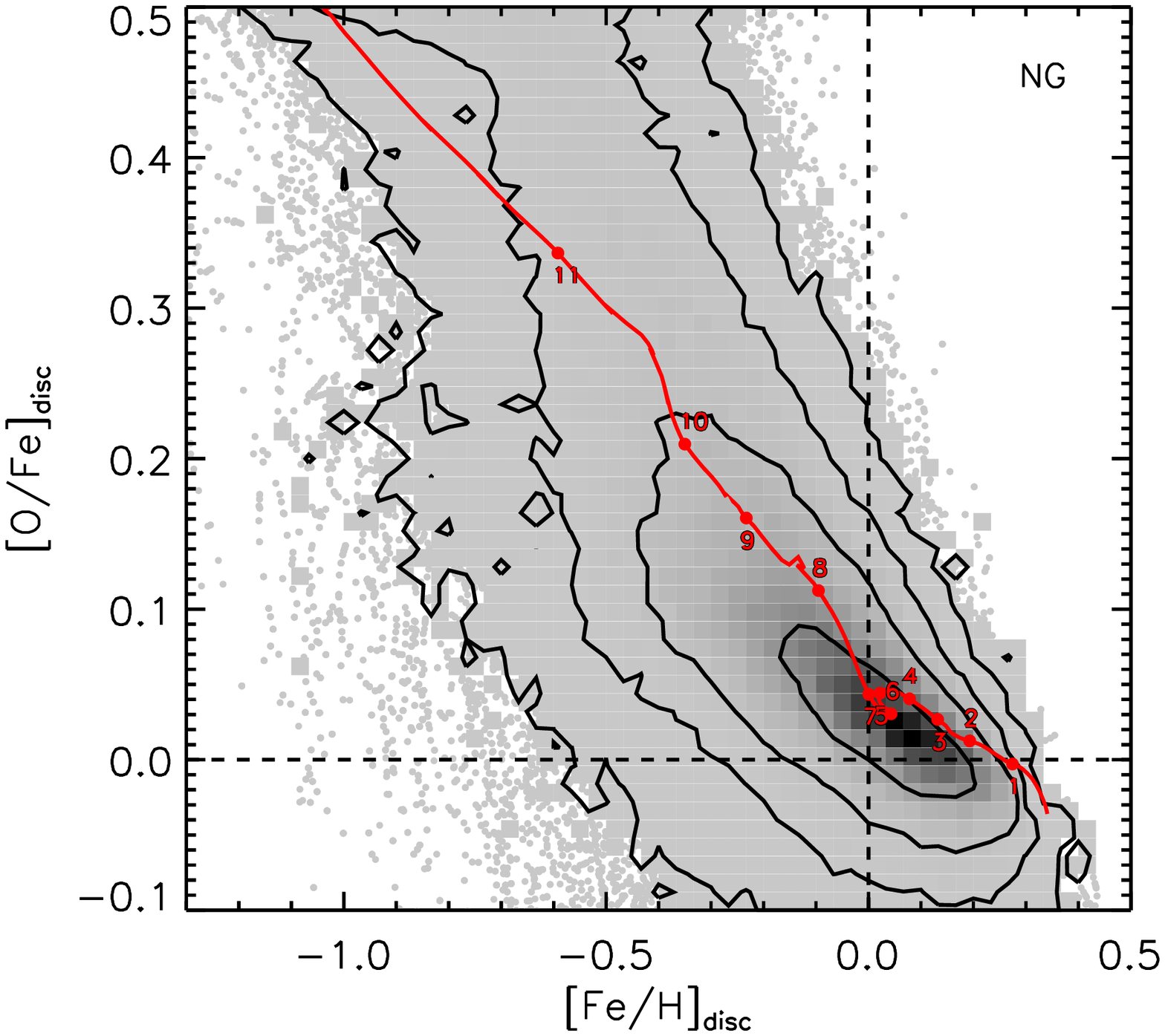} \\
\caption{The [Fe/H]-[O/Fe] relation for G dwarfs in the stellar discs of our MW-type model galaxy sample when using a bi-modal (top panel), power-law (middle panel), and narrow Gaussian (bottom panel) SN-Ia DTD. One galaxy contributes many hundreds of points to this relation (see \S \ref{sec:SFH, ZH and EH arrays}). The greyscale indicates the distribution of SSPs, weighted by the mass formed. Contours show the 68th, 95th, 99th and 99.9th percentiles. The chemical evolution of an individual MW-type model galaxy is over-plotted on each panel (red tracks), and discussed in detail in \S \ref{sec:An individual MW-type model galaxy}. Points on the track denote the chemical composition at discreet times in the past, labelled by the lookback time in Gyrs. The SFH of the same galaxy is plotted in red in Fig. \ref{fig:MW_SFHs}.}
\label{fig:FeH-OFe_SingleMW_AllDTDs}
\end{figure}

\subsection{The Milky Way disc} \label{sec:Milky Way stars}
There is now a wealth of data available in the literature on the chemical composition of stars in the MW disc. These data allow us to put firm constraints on the success of our GCE implementation in reproducing realistic MW-type model galaxies. We construct a sample of $\sim5200$ central (type 0) galaxies at $z=0$ that are disc dominated (i.e. $M_{\textnormal{bulge}}/(M_{\textnormal{bulge}}+M_{\textnormal{disc}}) < 0.5$), with DM halo masses in the range $11.5 \leq \textnormal{log}(M_{\textnormal{vir}})/\textnormal{M}_{\textnormal{\astrosun}} \leq 12.5$, and recent star formation rates of $1.0 \leq \textnormal{SFR}/\textnormal{M}_{\textnormal{\astrosun}}\textnormal{yr}^{-1} \leq 10.0$ over the redshift range $0.0 \leq z \leq 0.25$ (i.e. the last $\sim 3.0$ Gyrs). Our results are not affected by small changes to these criteria. Three example star formation histories (SFHs) from our MW-type model sample are shown in Fig. \ref{fig:MW_SFHs}. The chemical evolution of the individual galaxy depicted in red is discussed in \S \ref{sec:An individual MW-type model galaxy}. In this section, the model values are normalised to the solar abundances determined by \citet{AG89}.

\begin{figure*}
\centering
\begin{tabular}{@{}c@{} @{}c@{} @{}c@{}}
\includegraphics[totalheight=0.25\textheight, width=0.31\textwidth]{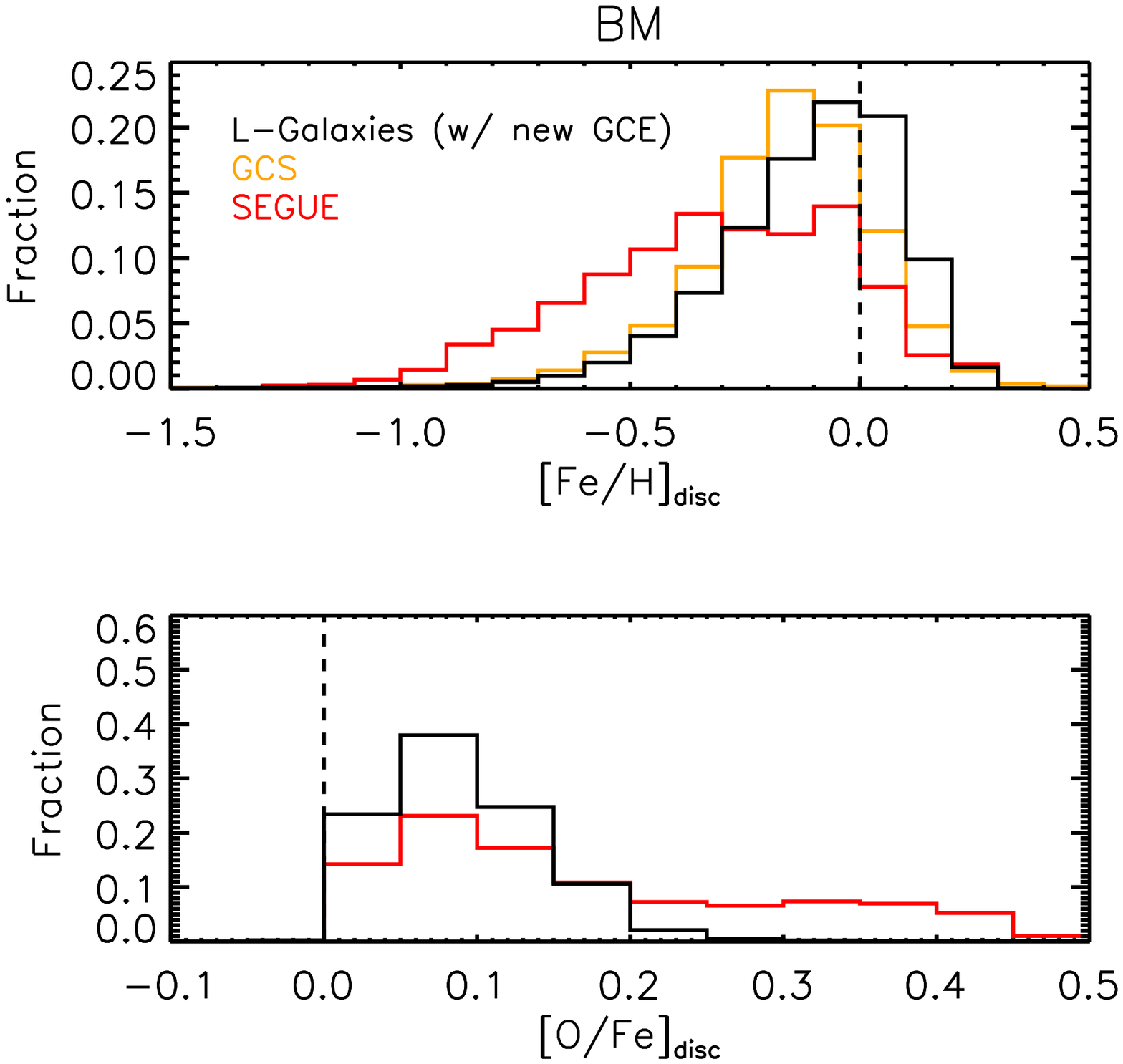} &
\includegraphics[totalheight=0.25\textheight, width=0.31\textwidth]{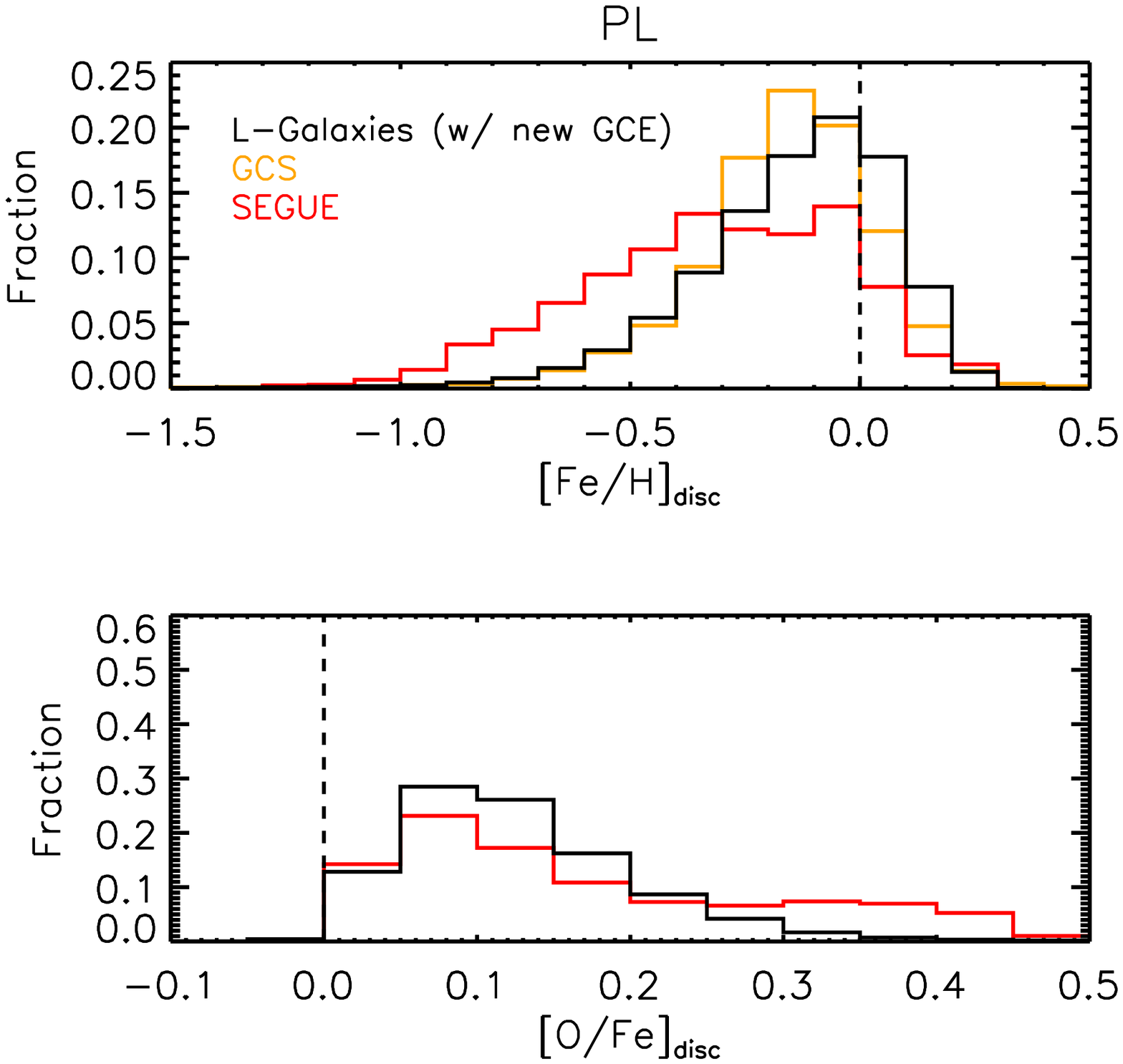} &
\includegraphics[totalheight=0.25\textheight, width=0.31\textwidth]{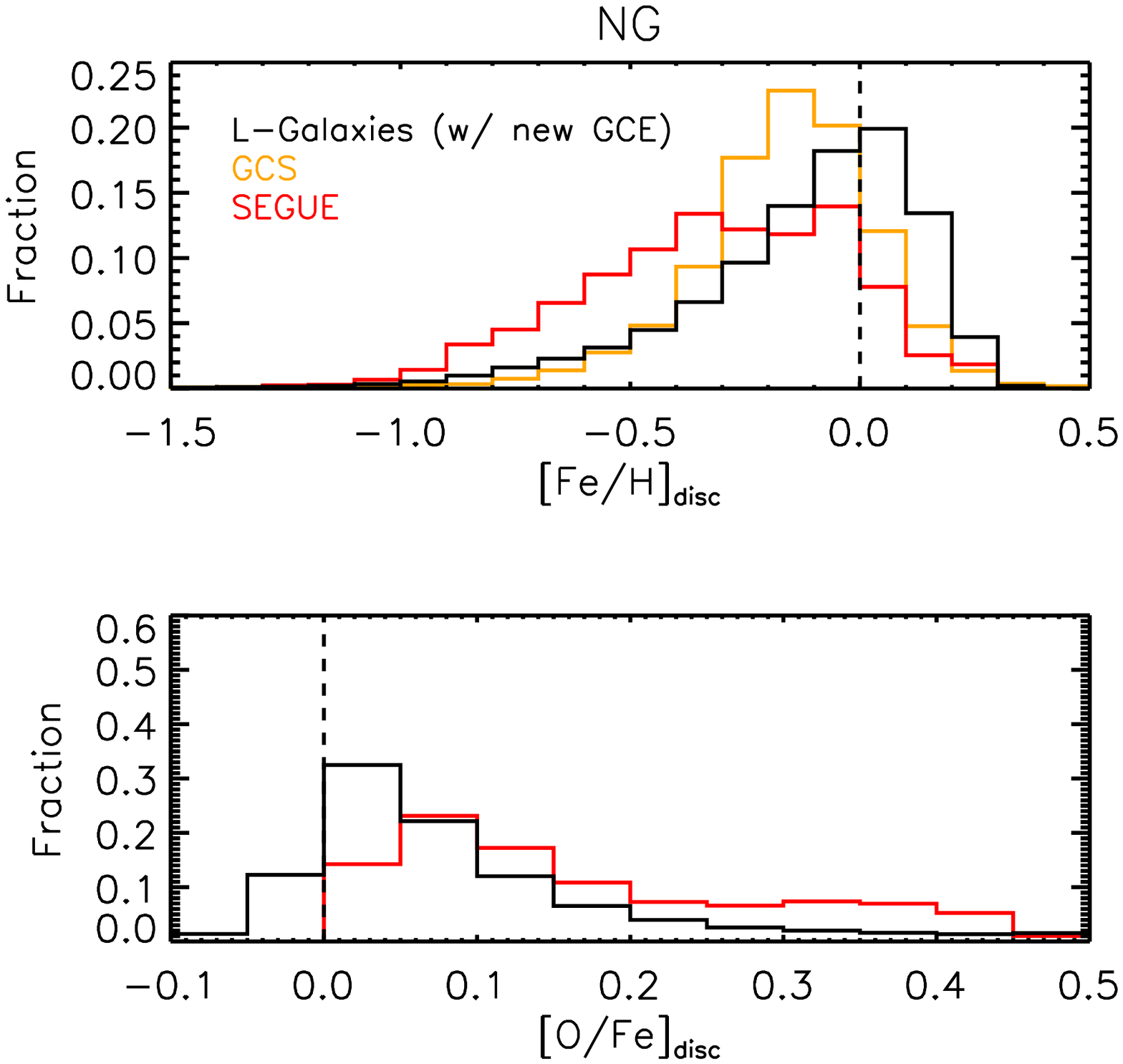} \\
\end{tabular}
\caption{\textit{Top row}: [Fe/H] distributions for the stellar discs of our model MW-type galaxies, when using a bi-modal (left), power-law (middle), or narrow Gaussian (right) SN-Ia DTD. Vertical dashed lines indicate the solar iron abundance. \textit{Bottom row}: [O/Fe] distributions for the same model discs and DTDs. Vertical dashed lines indicate the solar oxygen abundance.}
\label{fig:MW_FeH_and_OFe_dists_FP_ObsComp}
\end{figure*}

\begin{figure*}
\centering
\begin{tabular}{@{}c@{} @{}c@{} @{}c@{}}
\includegraphics[totalheight=0.25\textheight, width=0.31\textwidth]{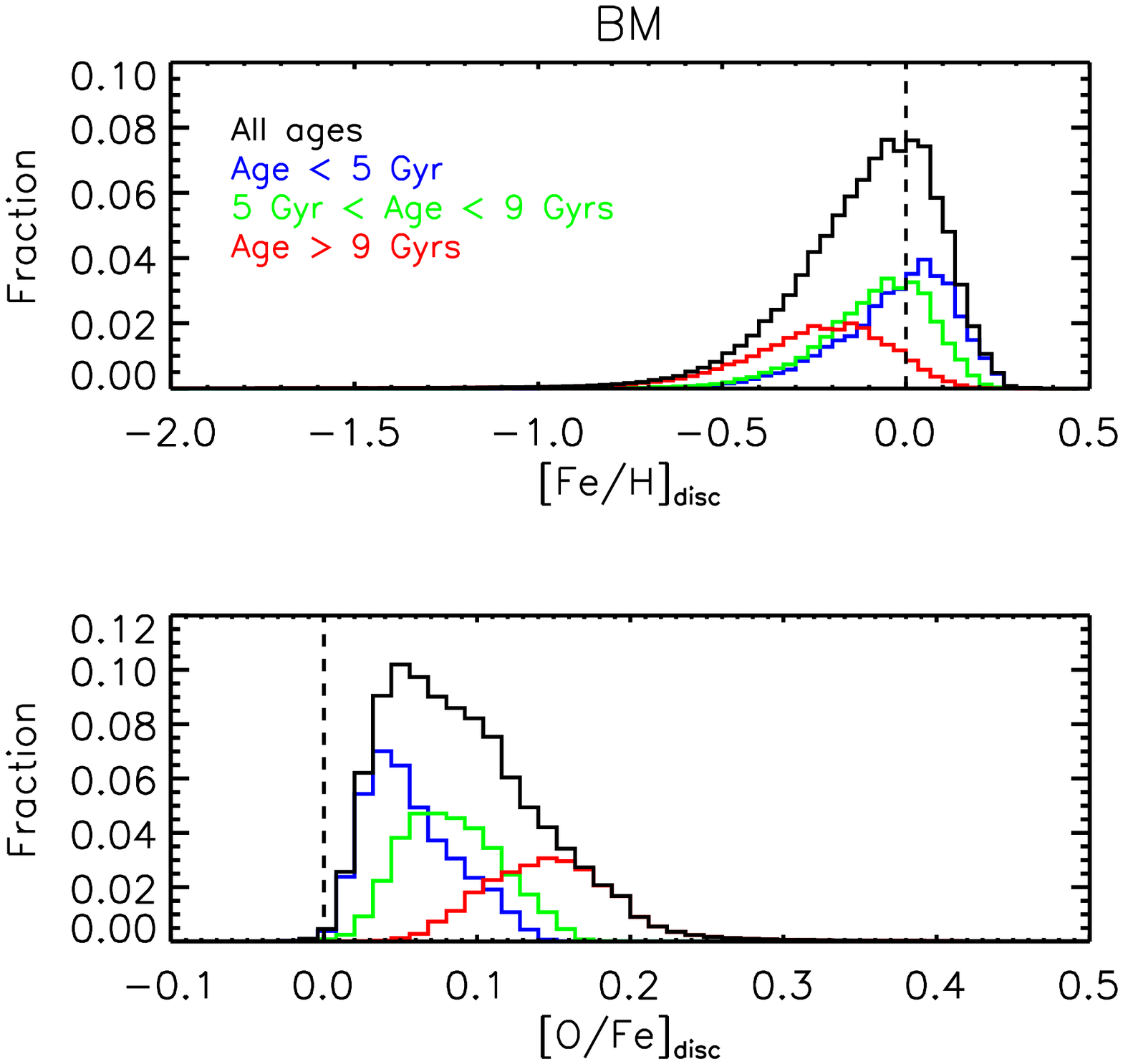} &
\includegraphics[totalheight=0.25\textheight, width=0.31\textwidth]{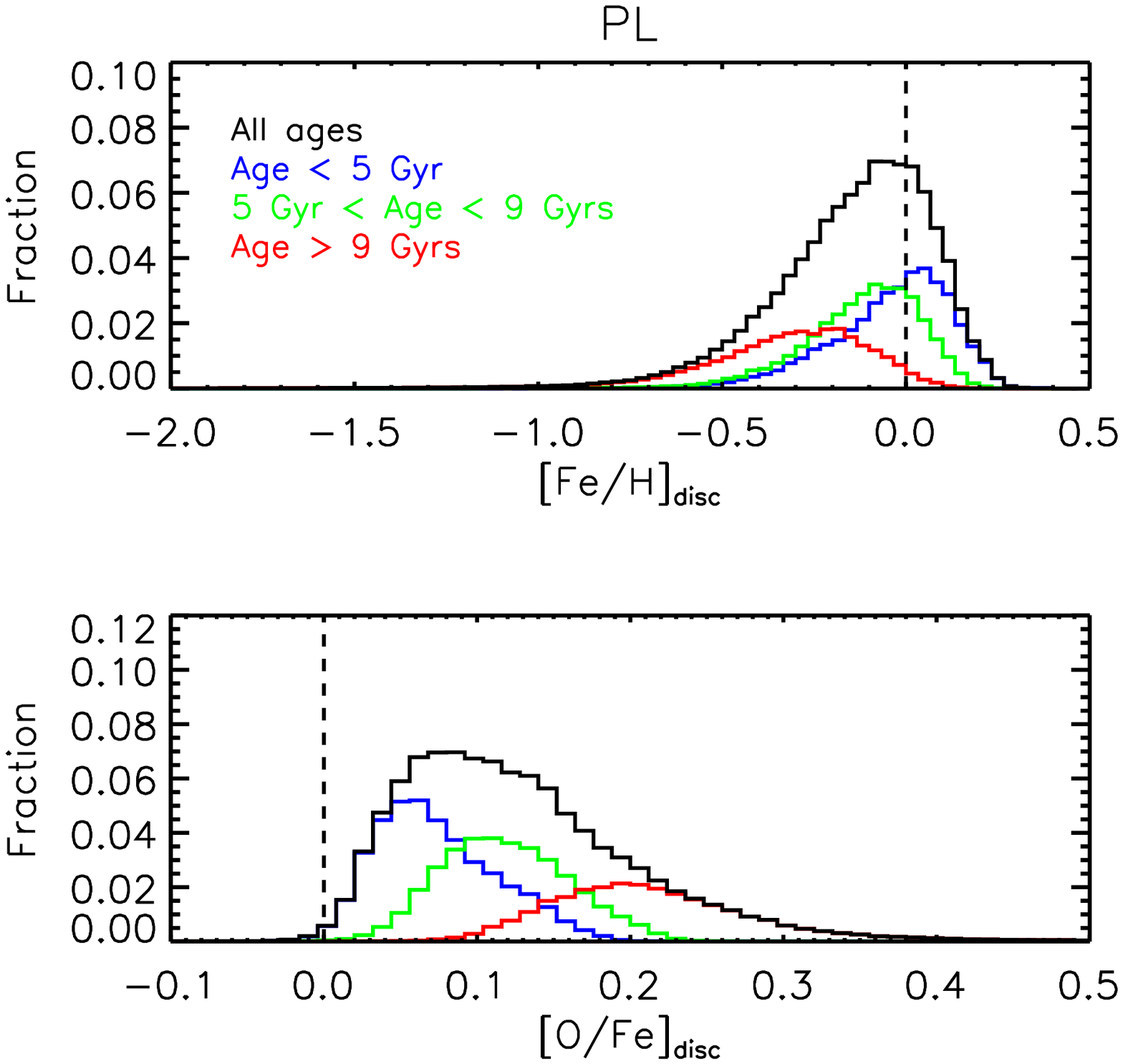} &
\includegraphics[totalheight=0.25\textheight, width=0.31\textwidth]{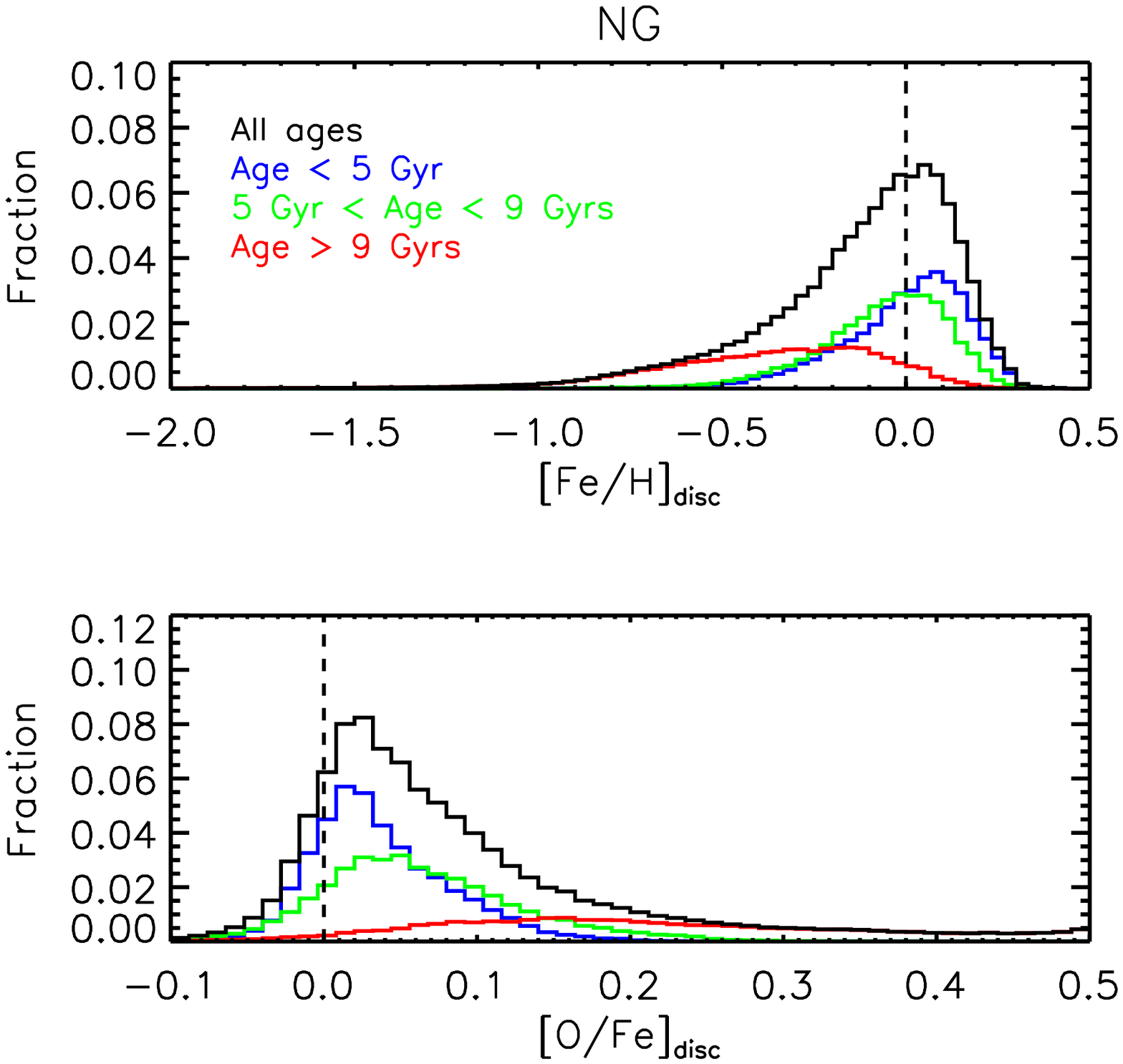} \\
\end{tabular}
\caption{\textit{Top row}: [Fe/H] distributions for the stellar discs of our model MW-type galaxies, when using a bi-modal (left), power-law (middle), or narrow Gaussian (right) SN-Ia DTD. Vertical dashed lines indicate the solar iron abundance. \textit{Bottom row}: [O/Fe] distributions for the same model discs and DTDs. Vertical dashed lines indicate the solar oxygen abundance.}
\label{fig:MW_FeH_and_OFe_dists}
\end{figure*}

In order to compare with observations, we only consider G dwarfs ($0.8 \leq M/\textnormal{M}_{\textnormal{\astrosun}} \leq 1.2$) still present in the stellar discs of our model MW-type galaxies at $z=0$. When using the P98 stellar lifetimes (\S \ref{sec:Stellar lifetimes}), \textit{not} all G dwarfs live as long as the age of the MW disc. For example, stars of mass $1.2 \textnormal{M}_{\textnormal{\astrosun}}$ (the upper mass limit we assume for G dwarfs) live from 3.1 Gyrs at $Z_{\textnormal{init}}=0.0004$ to a maximum of 4.7 Gyrs at $Z_{\textnormal{init}}=0.02$ (see Fig. \ref{fig:lifetimes}). These timescales are clearly shorter than the typical ages of the oldest SSPs in our MW-type model discs\footnote{The smallest G dwarfs considered ($0.8 \textnormal{M}_{\textnormal{\astrosun}}$) can live from around 14 to 26 Gyrs, and so \textit{do} survive the age of the disc.} (see Fig. \ref{fig:MW_SFHs}). Therefore, we re-weight those SSPs for which some of the G dwarfs would no longer be present at $z=0$ for the plots in this section. This correction removes a very small contribution from the oldest SSPs, reducing very slightly the number of low-[Fe/H], high-[O/Fe] stars. Although this is a more rigorous treatment, the main conclusions drawn from our MW-type sample also hold when assuming that all G dwarfs survive up to $z=0$.

\subsubsection{MW-type model galaxies} \label{sec:MW-type model galaxies}

Fig. \ref{fig:FeH-OFe_SingleMW_AllDTDs} shows the [Fe/H]-[O/Fe] relation for the G dwarfs in the stellar discs of our model MW-type galaxies, using the stitched-together histories described in \S \ref{sec:SFH, ZH and EH arrays}, for the three DTDs we consider. Care needs to be taken when comparing Fig. \ref{fig:FeH-OFe_SingleMW_AllDTDs} to observations. In observational studies of the MW disc, the chemical composition of individual stars of various ages are measured and plotted. In the case of our semi-analytic model, individual stars cannot be resolved, and so we must instead rely on the chemical composition of each \textit{population} of stars, formed at each timestep during the evolution of a galaxy. Fig. \ref{fig:FeH-OFe_SingleMW_AllDTDs} therefore shows the chemical composition of SSPs from $\sim 5200$ MW-type galaxies, where one MW-type galaxy contributes many hundreds of points (see \S \ref{sec:SFH, ZH and EH arrays}). Considering a whole sample of MW-type galaxies provides a statistically significant indication of the typical variation in the chemical composition of MW-type discs in our model. This method of comparison has been used before in semi-analytic models (e.g. \citealt{CM09}). Note that we therefore weight the SSPs by the mass of stars formed. The evolution of an example, individual MW-type galaxy is also plotted in each of the panels in Fig. \ref{fig:FeH-OFe_SingleMW_AllDTDs} (red tracks). This galaxy is discussed in detail in \S \ref{sec:An individual MW-type model galaxy}.

Each of the panels in Fig. \ref{fig:FeH-OFe_SingleMW_AllDTDs} shows a clear decrease in [O/Fe] with increasing [Fe/H] towards the solar composition. There are, however, important differences in the distribution of SSPs for each of the three DTDs we consider. These differences can be seen more clearly in Fig. \ref{fig:MW_FeH_and_OFe_dists_FP_ObsComp}, where we compare the [Fe/H] and [O/Fe] distributions when using our three DTDs (black histograms) with those of 16,134 F and G dwarfs from the \textit{Geneva-Copenhagen Survey} (\textit{GCS}, orange histograms) (\citealt{N04,HNA09}) and 293 unique G dwarfs from the \textit{Sloan Extension for Galactic Understanding and Exploration} (\textit{SEGUE}, red histograms) survey (\citealt{Y09,B12a,B12b}).

The difference in [Fe/H] distribution between the two observational samples is likely due to their different depths; the \textit{GCS} probed strictly the solar neighbourhood ($7.7 \lesssim R_{\textnormal{GC}}/\textnormal{kpc} \leq 8.31$ and $0.0 \leq |Z_{\textnormal{GC}}|/\textnormal{kpc} \leq 0.359$), whereas \textit{SEGUE} covered a wider range of galactic radii but also much higher galactic scale heights ($5 \lesssim R_{\textnormal{GC}}/\textnormal{kpc} \lesssim 12$ and $0.3 \leq |Z_{\textnormal{GC}}|/\textnormal{kpc} \leq 3.0$).\footnote{In Fig. \ref{fig:MW_FeH_and_OFe_dists_FP_ObsComp} the stars with $|Z_{\textnormal{GC}}|<0.3$ that are missing from the \textit{SEGUE} survey are accounted for via the mass re-weighting of the [Fe/H] distribution described by \citet{B12a}.} This means that the \textit{SEGUE} sample includes a larger number of metal-poor, `thick-disc' stars, and so has a [Fe/H] distribution spread to lower iron abundances. Our model, in turn, represents the \textit{average} chemical composition of stars born at each timestep in the discs of MW-type galaxies, due to the full mixing of material in the various galactic components.

\begin{figure*}
\centering
\includegraphics[totalheight=0.4\textheight, width=0.6\textwidth]{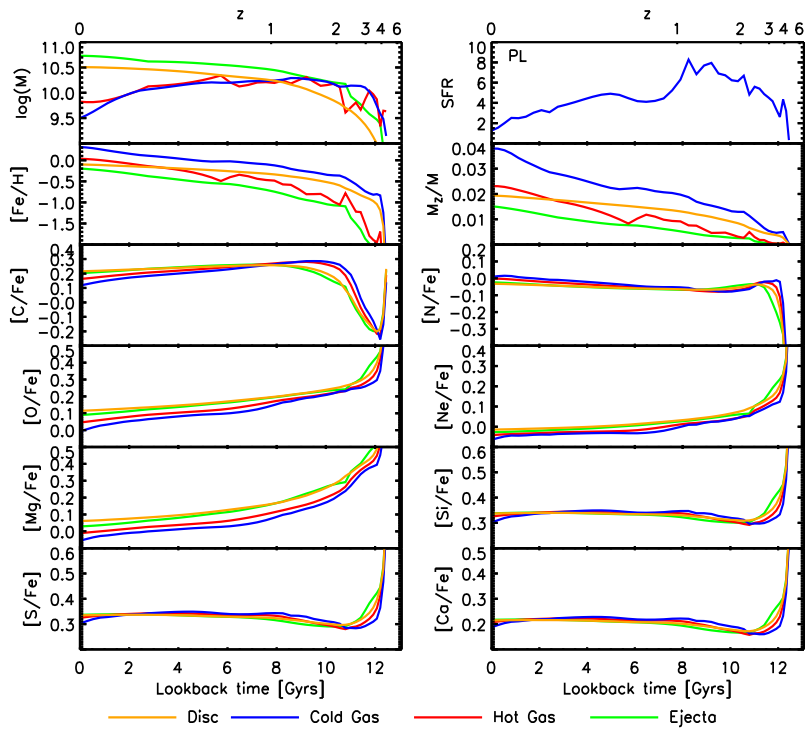}
\caption{The evolution, from redshift 7 to 0, of the mass (in M$_{\textnormal{\astrosun}}$), SFR (in M$_{\textnormal{\astrosun}}$/yr), iron abundance, total metallicity, and heavy element enhancements of four different galaxy components (see legend) in the example MW-type model galaxy shown in Fig. \ref{fig:FeH-OFe_SingleMW_AllDTDs}, using the power-law DTD.}
\label{fig:Evo_SingleMW}
\end{figure*}

The model [Fe/H] distributions for all three of our set-ups are in reasonable agreement with the \textit{GCS} data (partly by construction, as we have tuned $A$ to obtain a peak of the [Fe/H] distribution around 0.0), although the NG set-up is skewed slightly more to higher iron abundances. However, there are significant differences in the model [O/Fe] distributions. For example, the high-[O/Fe] tail in our BM set-up is much less extended than seen in the [$\alpha$/Fe] distribution from the \textit{SEGUE} survey\footnote{Note that \citet{B12a} and \citet{B12b} choose [$\alpha$/Fe] to be the average of [Mg/Fe], [Si/Fe], [Ca/Fe] and [Ti/Fe], with no oxygen lines included in the analysis. However, as oxygen is the most abundant $\alpha$ element in galaxies, a comparison between their [$\alpha$/Fe] and our [O/Fe] is still valid here.}. This suggests that stars are being enriched with iron too quickly when using the bi-modal DTD -- a conclusion also reached by \citet{M09}.

Interestingly, the extent of the high-[O/Fe] tail in the [O/Fe] distribution increases from left to right in Fig. \ref{fig:MW_FeH_and_OFe_dists_FP_ObsComp}. This is due to the different number of `prompt' SNe-Ia assumed for each of the three DTDs. The smaller the prompt component, the larger the number of low-[Fe/H], high-[O/Fe] stars that can be formed before a significant amount of Fe gets into the star-forming gas. The bi-modal DTD allows $\sim 54$ per cent of SNe-Ia to explode within 100 Myrs of star formation ($\sim 58$ per cent within 400 Myrs), the power-law DTD allows $\sim 23$ per cent within 100 Myrs of star formation ($\sim 48$ per cent within 400 Myrs), and the Gaussian has no prompt component at all. Only the Gaussian DTD has a high-[O/Fe] tail as extended as that seen for G dwarfs from \textit{SEGUE}. However, we reiterate that the lack of any prompt component is in contradiction with recent observations (e.g. \citealt{MM12}). The smaller high-[O/Fe] tail produced when using the power-law DTD, although not as extended as seen in the \textit{SEGUE} data, is still promising, espcially when considering that a) the \textit{SEGUE} data contain a large number of $\alpha$-enhanced, iron-poor stars at high galactic scale heights, and b) our model represents the chemical composition of MW-type stellar discs in a statistical sense, and also assumes full mixing of metals in the stellar disc.\footnote{Including a treatment of the radial distribution of metals in galaxies, similar to that done by \citet{F13}, will be the focus of future work.} We will also show in \S \ref{sec:Ellipticals} that the power-law DTD also produces positive slopes in the $M_{*}$-[$\alpha$/Fe] relations of elliptical galaxies.

In Fig. \ref{fig:MW_FeH_and_OFe_dists} we show a finer binning of the model [Fe/H] and [O/Fe] distributions for the three SN-Ia DTDs considered (black histograms). Sub-distributions for three distinct age ranges (coloured histograms) are also plotted. All panels show nicely that older SSPs have lower [Fe/H] and higher [O/Fe] than younger SSPs, due to the delayed enrichment of the star-forming gas with iron from SNe-Ia. There is also no sign of an extended tail below $\textnormal{[Fe/H]}=-1.0$ (the `G-dwarf problem') that is common to closed-box models.

The broader plateau present in the [O/Fe] distribution for the PL set-up is due to the shape of the DTD; the power-law DTD assumes a smoother change in SN-Ia rate with time than the other two DTDs considered (see Fig. \ref{fig:DTDs}). This means that the ISM in a typical MW-type galaxy undergoes a fairly constant decrease in [O/Fe] of $\sim 0.025$ dex/Gyr for the power-law DTD. In contrast, the bi-modal DTD causes a more gradual decrease in [O/Fe] of $\sim 0.016$ dex/Gyr, \textit{after} significantly enriching the ISM with iron shortly after the start of star formation. In turn, the Gaussian DTD produces a steep decrease in [O/Fe] of $\sim 0.066$ dex/Gyr from very high initial values until $\sim 1$ Gyr after the peak of star formation, with little change thereafter.

The CL04 SN-II yields produce qualitatively similar results to those discussed above, except that the [Fe/H] distribution is shifted to higher values and the [O/Fe] has a decreased high-[O/Fe] tail -- in greater contradiction with observations. This is because, when extrapolated to $120 \textnormal{M}_{\textnormal{\astrosun}}$, the CL04 yields predict a higher production of O, Mg and Fe by SNe-II than the P98 yields, particularly at low metallicity.

\subsubsection{An individual MW-type model galaxy} \label{sec:An individual MW-type model galaxy}
In this sub-section, we look more closely at the chemical evolution of an individual MW-type galaxy in our model. This galaxy's SFH is plotted in red in Fig. \ref{fig:MW_SFHs}, and its evolution in the [Fe/H]-[O/Fe] diagram is shown by a red track in each panel of Fig. \ref{fig:FeH-OFe_SingleMW_AllDTDs}. Points on the tracks in Fig. \ref{fig:FeH-OFe_SingleMW_AllDTDs} denote the chemical composition at discreet times in the past, labelled by the lookback time in Gyrs.

This galaxy nicely demonstrates the fairly smooth evolution that we would expect from a MW-type galaxy. However, it is not necessarily typical of our MW-type model sample as a whole. Some galaxies (such as that shown in blue in Fig. \ref{fig:MW_SFHs}) undergo large infall and star formation events that can cause such a track to double-back on itself and otherwise deviate from a `smooth' path (see also \citealt{CM09}). However, our chosen galaxy provides a good example of the general chemical evolution undergone by MW-type galaxies in our model.

Fig. \ref{fig:Evo_SingleMW} shows the evolution from $z=7$ to 0 of the mass, SFR, iron abundance, total metallicity, and complete set of heavy element enhancements for this example MW-type galaxy, when using the power-law DTD. The different components of the galaxy (stellar disc, cold gas, hot gas, and ejecta reservoir) are coloured according to the legend.\footnote{For clarity, the bulge component is not plotted in Fig. \ref{fig:Evo_SingleMW}. A small bulge of $4.8\times 10^{8} M_{\textnormal{\astrosun}}$ is formed via a very minor merger (68:1 ratio) in this galaxy at $z\sim 8.5$, without any accompanying disturbance of the stellar disc. The bulge inherited the chemical composition of the satellite's stars at that time.} Note that Fig. \ref{fig:Evo_SingleMW} shows the \textit{average} chemical composition of a whole galaxy component at any given time.

Fig. \ref{fig:Evo_SingleMW} highlights the dependence of an element's evolution on the mode of its release, namely SNe-II, SNe-Ia or AGB winds. Those elements that are predominantly produced in massive SNe-II (O, Ne and Mg) show a similar decline in their enhancement with cosmic time, as we would expect for a slowly declining SFR and a delayed enrichment of iron. These light $\alpha$ elements also show lower enhancements in the cold gas than in the stellar disc for this reason. The heavier $\alpha$ elements (Si, S and Ca) are produced mainly in lower-mass SNe-II, and also have a greater contribution from SNe-Ia. They are therefore released into the ISM later than the lighter $\alpha$ elements, showing a gradual increase in enhancement with time (at a decreasing rate), and higher enhancements in the gas than in the stars while gas fractions are high. Nitrogen, an element with a dominant contribution from (delayed) AGB winds at low metallicity, shows a strong increase in [N/Fe] at the onset of AGB wind enrichment (at $z\sim 4$ for this galaxy), followed by a more gradual increase thereafter. Finally, the drop in [C/Fe] between $z\sim 3$ and 4 is due to a decrease in the C/Fe ratio in the ejecta of SNe-II at $Z_{\textnormal{init}}\sim 0.004$ compared to other metallicities. The increase in this ratio at higher metallicities, along with a significant contribution to C from AGB winds, casues the sharp rise in [C/Fe] shortly after. This is a specific property of the P98 SN-II yields. When using the CL04 SN-II yields, the [C/Fe] evolution follows that of the light $\alpha$ elements more closely.

\textit{} \newline
To conclude this section, we can say that our new GCE implementation is able to reproduce the [O/Fe] distribution for G dwarfs in the MW disc \textit{if} there is only a \textit{minor} prompt component of SNe-Ia (i.e. $\leq 50$ per cent within $\sim 400$ Myrs). Our NG set-up (narrow Gaussian DTD, delayed SNe-Ia only) and PL set-up (power-law DTD, $\leq 48$ per cent of SNe-Ia exploding within $\sim 400$ Myrs) therefore reproduce the observed high-[O/Fe] tail best. However, the power-law DTD achieves this whilst assuming a more realistic fraction of prompt SNe-Ia. Our new GCE implementation also allows us to examine, in detail, the chemical evolution undergone by individual galaxies, which can help us explain the features seen in the sample as a whole.

\subsection{Elliptical galaxies} \label{sec:Ellipticals}
The change in various element ratios as a function of velocity dispersion or $M_{*}$ in ellipticals can also provide insight into the chemical evolution of galaxies. It has been observed that $\alpha$ enhancements increase with $M_{*}$ (e.g. \citealt{GFS09,T10,JTM12,CGvD13}). This has been mainly attributed to massive ellipticals undergoing the majority of their star formation at higher redshifts and over shorter timescales. The stars in these galaxies are therefore likely to be deficient in iron, as they were formed before a significant number of SNe-Ia could enrich the star-forming gas. Less massive ellipticals, on the other hand, are believed to form a larger fraction of their stars later, from gas that has had time to be more enriched with iron. These galaxies should therefore have lower stellar $\alpha$ enhancements.

\begin{figure}
\centering
\includegraphics[totalheight=0.2\textheight, width=0.46\textwidth]{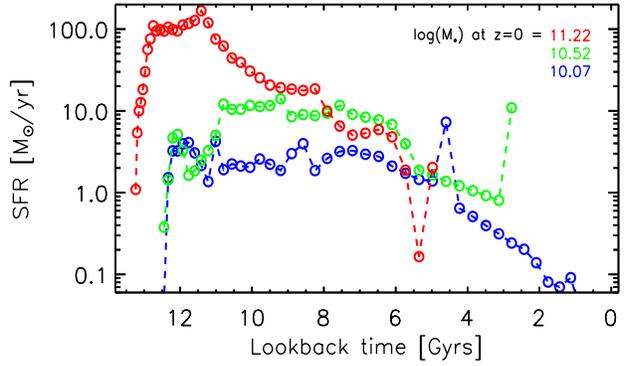}
\caption{Three example SFHs from our model elliptical sample. The different colours correspond to different stellar masses at $z=0$ (see legend). The points represent the sum of the SFRs from all progenitors at every output snapshot of the simulation. Low-mass ellipticals tend to have longer star-formation timescales than high-mass ellipticals in our model.}
\label{fig:Ellip_SFHs}
\end{figure}

Previous GCE models, working within a hierarchical merging scenario, have found it difficult to reproduce this trend between stellar mass and $\alpha$ enhancement, without invoking either a variable or adapted IMF, morphologically-dependent star formation efficiencies (SFEs), or additional prescriptions to increase star formation at high redshift (e.g. \citealt{TGB99,T99,N05b,P09b,CM09,A10a,A10b,CM11}).

We select $z=0$ elliptical galaxies by bulge-to-mass ratio and (g-r) colour, such that $M_{\textnormal{bulge}}/(M_{\textnormal{bulge}}+M_{\textnormal{disc}}) \geq 0.7$ and (g-r) $\geq 0.051\:\textnormal{log}(M_{*}) + 0.14$, to form a sample of $\sim 8700$ galaxies. These cuts are chosen to match the selection criteria used to obtain the sample of SDSS-DR7 ellipticals shown as green points in Fig. \ref{fig:Ellip_XFe_Mstar}. The (g-r) colour cut also nicely separates the red sequence from the blue cloud in our model at $z=0$. Our model sample includes type 0, 1 and 2 galaxies (see \S \ref{sec:Infall, Cooling and Outflows}).

Fig. \ref{fig:Ellip_SFHs} shows the SFHs of three example galaxies from our model elliptical sample. In this case, the sum of the SFRs from all progenitors at any given snapshot are plotted, rather than the SFRs from only the main progenitors. We can see that the lowest-mass elliptical (blue) has a more extended SFH than the highest-mass elliptical (red). This is the case in general for our model elliptical sample (see also \citealt{DL06}). We can also see that the two most massive ellipticals in Fig. \ref{fig:Ellip_SFHs} (red and green) had their star formation shut-down after a merger-induced starburst at $\sim 5$ and $\sim 3$ Gyrs lookback time, respectively.

\begin{figure}
\centering
\includegraphics[totalheight=0.32\textheight, width=0.46\textwidth]{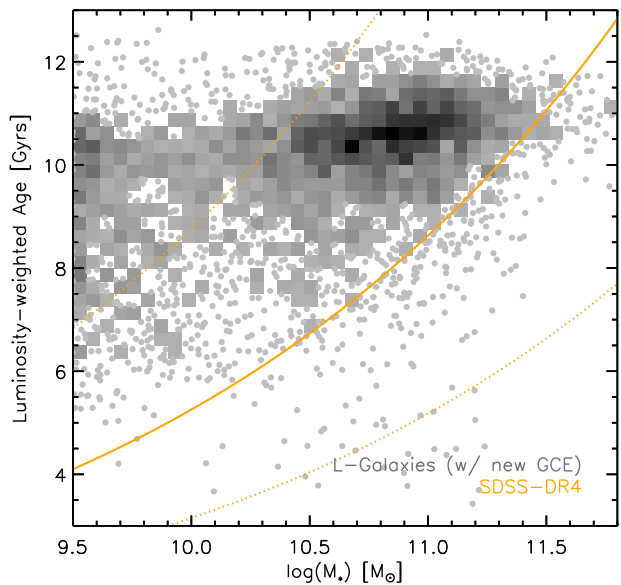}
\caption{The $M_{*}$-age relation for our model elliptical galaxies. Greyscale denotes the number density of galaxies. Model ages are weighted by their r-band luminosity. A linear fit to the same relation for the SDSS-DR4 from JTM12 is given by the solid orange line, with the 1$\sigma$ spread given by dotted orange lines. Our mass-age-selected sub-sample is made up of model galaxies that lie within one standard deviation ($\pm0.222$ dex) of the observed mass-age relation.}
\label{fig:Ellip_Mstar_Age}
\end{figure}

\subsubsection{The mass-age relation}
Before discussing element enhancements, we first show the $M_{*}$-age relation for our model elliptical sample in Fig. \ref{fig:Ellip_Mstar_Age}. Also shown is a fit to the luminosity-weighted mass-age relation from the \Citeauthor{JTM12} (2012, hereafter JTM12) sample (solid orange line) and its $1\sigma$ scatter (dotted orange lines). The ages of model galaxies are r-band luminosity weighted in this plot, in order to make a fairer comparison with the observations. For the observed relation, we have used the stellar masses taken directly from the SDSS-DR7 catalogue\footnote{Available at http://www.mpa-garching.mpg.de/SDSS/DR7}. This is also the case for all subsequent plots showing data from the JTM12 sample.

It is known that semi-analytic models tend to produce too many old, red, dwarf galaxies by $z=0$ compared to observations (e.g. \citealt{W06,G10}), as can be seen by comparing the model and observations in Fig. \ref{fig:Ellip_Mstar_Age}. This is caused not just by the strong stripping of gas from satellites, but also by the strong SN feedback required to match the observed galaxy stellar mass function. Recent work by \citet{H13} has improved this problem to some extent, by allowing material ejected from model galaxies at high-$z$ to be reaccreted over longer timescales, allowing them to form more stars at low-$z$, and therefore be younger and bluer at $z=0$. However, this improvement is not implemented into the \textsc{L-Galaxies} model presented here. Therefore, in the following sections, we will distinguish between our full elliptical model sample and a `mass-age-selected' sub-sample, which includes only those model galaxies that lie within the 1$\sigma$ scatter of the observed $M_{*}$-age relation. This is \textit{not} done in order to evade the evident issues still affecting the galaxy formation model, but rather as a means of testing the relation between mass, age and $\alpha$ enhancement in our new GCE implementation.

\subsubsection{[$\alpha$/Fe] relations} \label{sec:[a/Fe] relations}
\begin{figure}
\centering
\includegraphics[totalheight=0.2\textheight, width=0.46\textwidth]{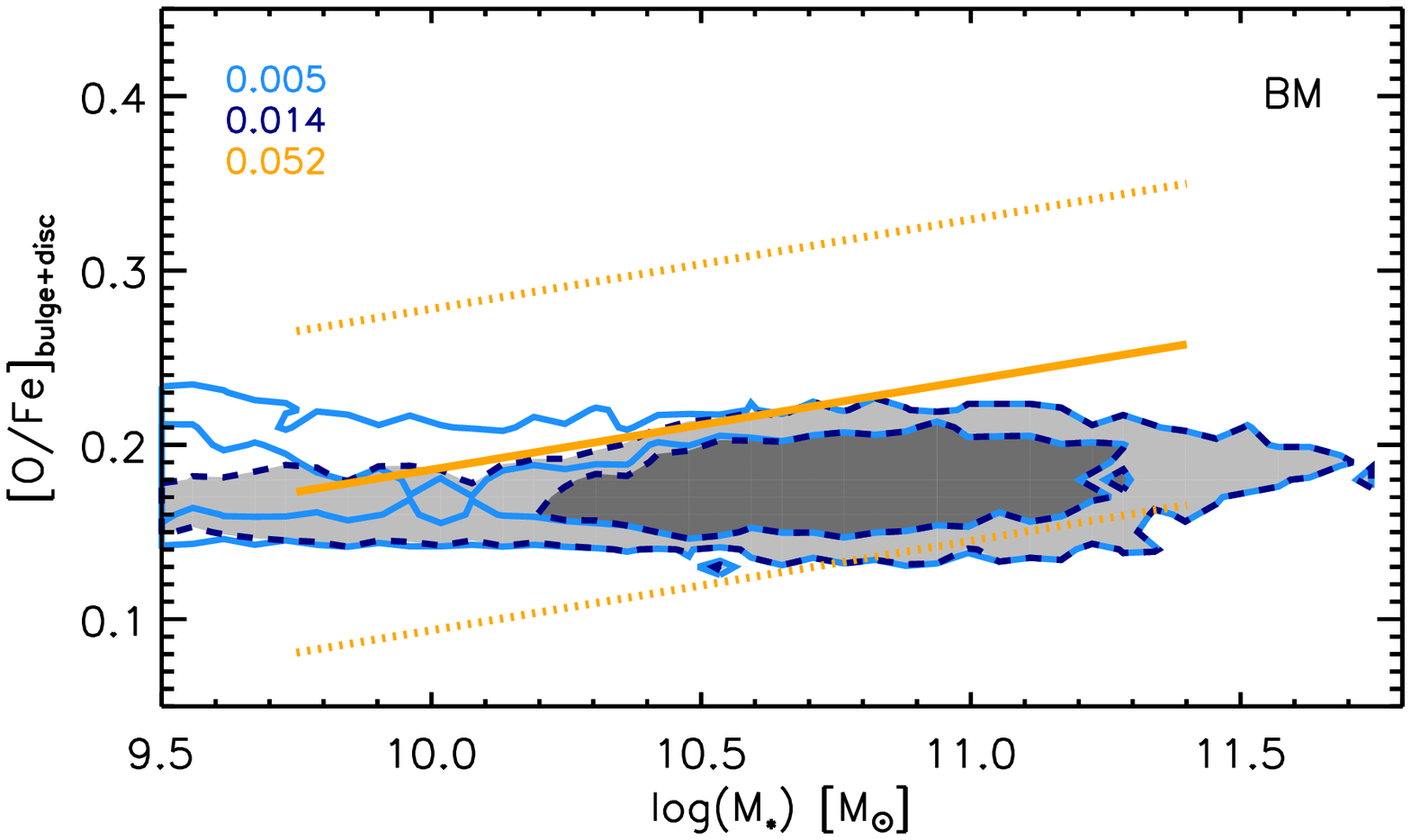} \\
\includegraphics[totalheight=0.2\textheight, width=0.46\textwidth]{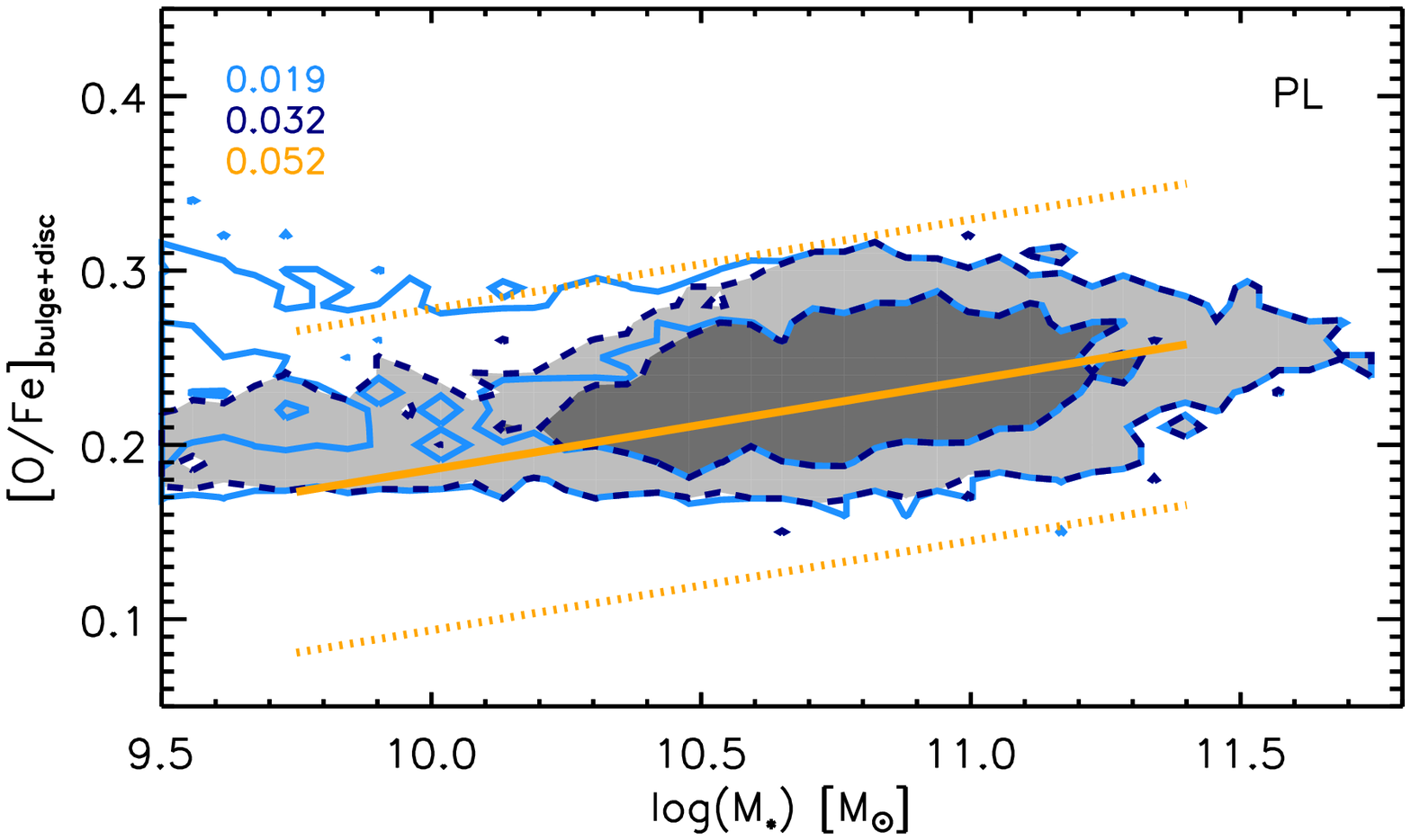} \\
\includegraphics[totalheight=0.2\textheight, width=0.46\textwidth]{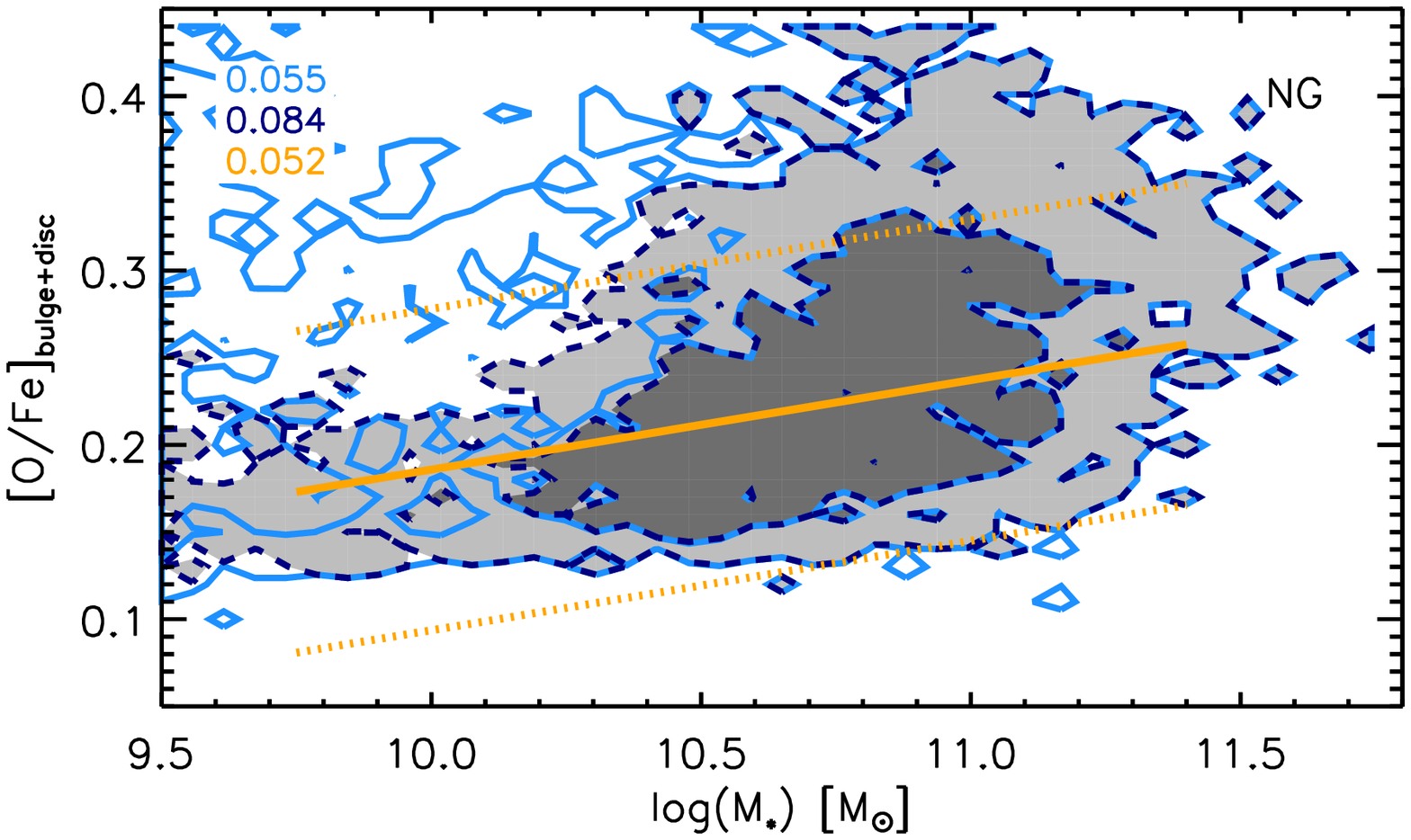} \\
\caption{The $M_{*}$-[O/Fe] relation for the bulge and disc components of our model elliptical sample, when using a bi-modal (top panel), power-law (middle panel), or narrow Gaussian (bottom panel) SN-Ia DTD. Light-blue contours represent our full elliptical sample. Dark-blue, dashed, filled contours represent out mass-age-selected sub-sample (see text and Fig. \ref{fig:Ellip_Mstar_Age}). Contours represent the 68th and 95th percentiles. A linear fit to the observed relation from JTM12 is given by the orange line, with its 1$\sigma$ spread (dotted orange lines). The slopes of these three relations are given in the top left corner of each panel.}
\label{fig:Ellip_OFe_Mstar_allDTDs}
\end{figure}

Fig. \ref{fig:Ellip_OFe_Mstar_allDTDs} shows the $M_{*}$-[O/Fe] relation for the bulge and disc stars of our model ellipticals at $z=0$, for the three SN-Ia DTDs we consider. Light-blue contours represent our full elliptical sample. Dark-blue, dashed, filled contours represent out mass-age-selected sub-sample. The observed relation from the JTM12 sample is given by the solid orange line. The slopes of the linear fits to these three relations are given in the top left corner of each panel.\footnote{The model slopes have been obtained from a linear fit in the range $10.0 \leq \textnormal{log}(M_{*}/\textnormal{M}_{\textnormal{\astrosun}}) \leq 12.0$.} Model element ratios in this section have been normalised to the solar abundances measured by \citet{GNS96}, in accordance with the observations to which we compare.

\begin{figure}
\centering
\includegraphics[totalheight=0.32\textheight, width=0.46\textwidth]{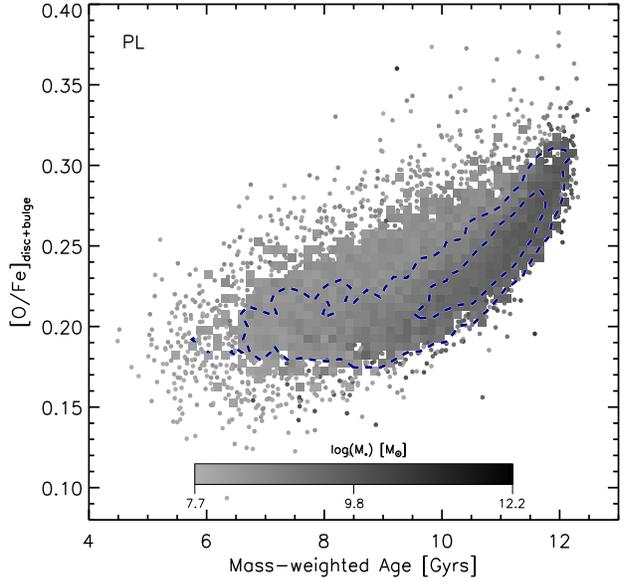}
\caption{The relation between mass-weighted age and oxygen enhancement for our model elliptical galaxies, when using the power-law DTD. Points show the full elliptical sample, with the greyscale indicating the stellar mass. Dark blue, dashed contours show the mass-age-selected sub-sample. There is a clear positive correlation between age and [O/Fe] in our model ellipticals, both in general and at fixed mass.}
\label{fig:Ellip_Age_OFe}
\end{figure}

We note here that estimates of element enhancements from stellar population synthesis (SPS) models, such as those used by JTM12, are found to be fairly good representations of the true global value, and are not as biased by small younger populations as age estimates can be \citep{ST07}. It is therefore reasonable for us to compare our model mass-weighted element enhancements with these observations.

As with the extent of the high-[O/Fe] tail in our MW-type sample (see \S \ref{sec:MW-type model galaxies}), we can see from Fig. \ref{fig:Ellip_OFe_Mstar_allDTDs} that the strength of the slope in the model $M_{*}$-[O/Fe] relation is inversely proportional to the fraction of prompt SNe-Ia assumed. For the BM set-up ($\sim 54$ per cent of SNe-Ia explode within 100 Myrs, $\sim 58$ per cent within 400 Myrs), the model slope is much flatter than observed. For the PL set-up ($\sim 23$ per cent of SNe-Ia explode within 100 Myrs, $\sim 48$ per cent within 400 Myrs), a positive slope is obtained, although shallower than observed. For the NG set-up (with no prompt component), a strong slope is obtained, with a larger scatter. 

The increase in slope is because massive ellipticals have shorter star-formation timescales in the model, and so decreasing the fraction of prompt SNe-Ia has a bigger effect on their final iron abundance, increasing their final $\alpha$ enhancements more than for low-mass ellipticals. The increase in scatter at fixed mass is because older galaxies have shorter star-formation timescales in the model, and so undergo a greater increase in their $\alpha$ enhancements for the same reason. This correlation between mass, age and $\alpha$ enhancement can also be seen in the increasing difference between the slope for the full elliptical sample and the mass-age-selected sub-sample from the top to bottom panel, and also from the age-[O/Fe] relation shown in Fig. \ref{fig:Ellip_Age_OFe} for the PL set-up. This result supports the canonical thinking that the slopes in $M_{*}$-[$\alpha$/Fe] relations are driven by differences in the star-formation timescale. If correct, then our model suggests there should only be a minor fraction of prompt SNe-Ia for any given SSP (i.e. $\leq 50$ per cent within $\sim 400$ Myrs).

\begin{figure*}
\centering
\includegraphics[totalheight=0.6\textheight, width=0.95\textwidth]{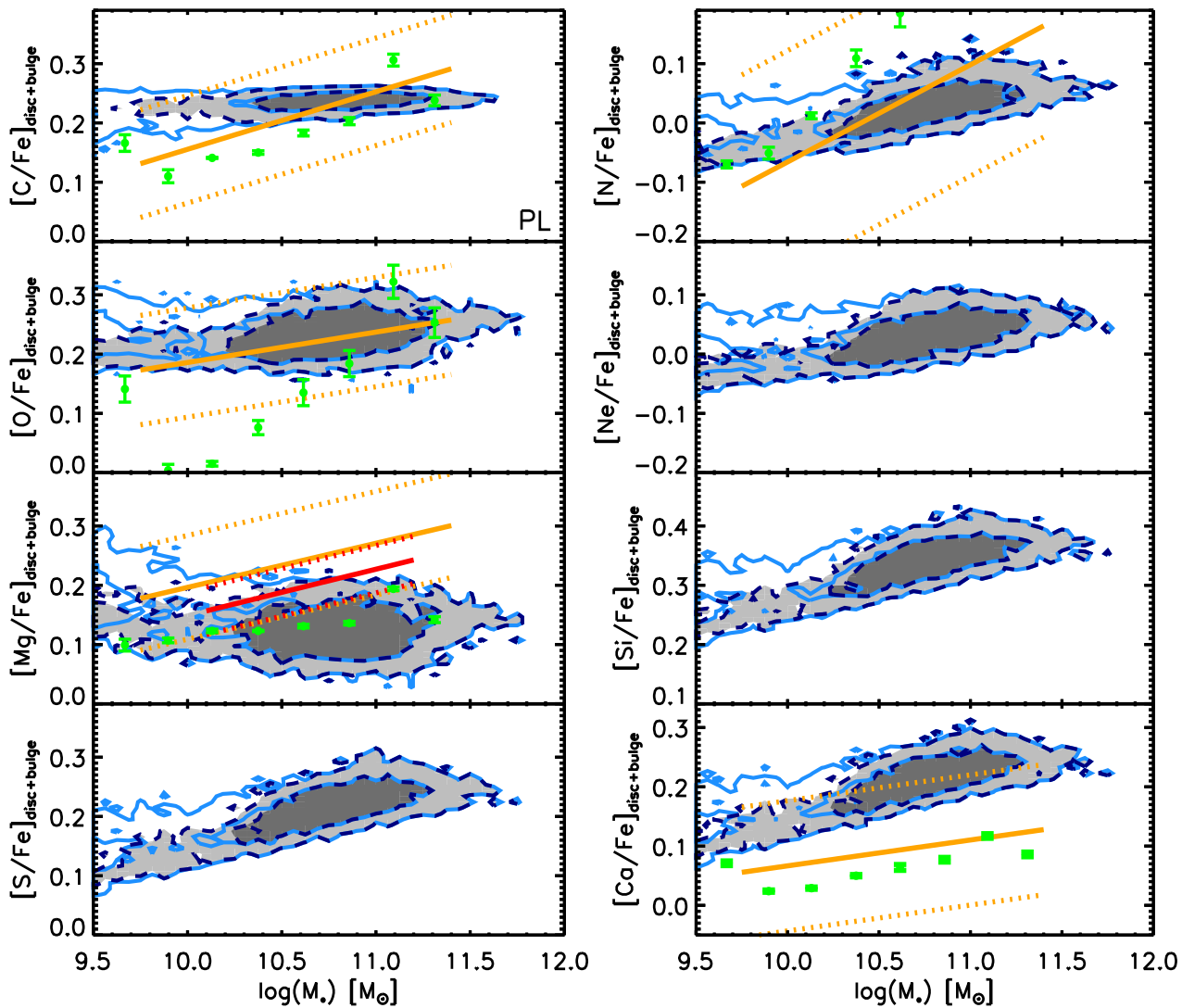}
\caption{Element enhancements as a function of stellar mass for the bulge and disc components of our model elliptical sample for our PL set-up. Light-blue contours represent our full elliptical sample. Dark-blue, dashed, filled contours represent out mass-age-selected sub-sample. For both samples, the contours show the 68th and 95th percentiles. Fits to the observed relations from the JTM12 sample (solid orange lines, with 1$\sigma$ scatter given by dotted orange lines), Graves et al. (2009) (solid red lines, with 1$\sigma$ scatter given by dotted red lines) and a newly-selected SDSS-DR7 sample (C. Conroy \& G. Graves, priv. comm.) (green points) are also shown for comparison.}
\label{fig:Ellip_XFe_Mstar}
\end{figure*}

Fig. \ref{fig:Ellip_XFe_Mstar} shows the enhancements of all the heavy elements that we track as a function of $M_{*}$ when using the power-law DTD. As in Fig. \ref{fig:Ellip_OFe_Mstar_allDTDs}, light-blue contours represent our full elliptical sample. Dark-blue, dashed, filled contours represent out mass-age-selected sub-sample. Fits to observations of ellipticals drawn from the SDSS-DR4 (orange lines, JTM12), the SDSS-DR6 (red line, \citealt{GFS09}), and the SDSS-DR7 (green points, see below) are also shown where possible. The PL set-up is shown here because it provides a positive slope for the $M_{*}$-[O/Fe] relation, while also assuming a more realistic fraction of `prompt' SNe-Ia than the NG set-up.

Pleasingly, Fig. \ref{fig:Ellip_XFe_Mstar} shows that positive slopes are obtained for all the $\alpha$ elements when using our PL set-up (except for Mg, as explained below). This is again because of the correlation between mass, age and $\alpha$ enhancement in our model. The same is true for our NG set-up, but not for the BM set-up, which has a large fraction of prompt SNe-Ia.  This is an important result, as it has been difficult previously for models to obtain positive slopes without invoking additional physics. It should be noted that the slopes of the different observational data shown in Fig. \ref{fig:Ellip_XFe_Mstar} differ substantially for some element enhancements. This is mainly due to the difference in the SPS modelling techniques used. Therefore, it is more important that our model produces positive slopes at all than reproduces the exact slopes of any particular observational data set. 

The methodology of both JTM12 and \citet{GFS09} is based on fitting observed and modelled Lick absorption line indices (e.g. \citealt{W94}). JTM12 adopt the SPS models of \citet{TMJ11b} and 18 Lick indices, whereas \citet{GFS09} adopt the models of \citet{S07} and use 7 Lick indices. The fits from JTM12 are based on a sample of visually-classified, early-type galaxies in the redshift range $0.05<z<0.06$. \citet{GFS09} selected red-sequence galaxies, classified by the colour-magnitude diagram, in the redshift range $0.04<z<0.08$. Both samples exclude star-forming galaxies by applying cuts to certain emission line strengths. Stellar masses are obtained from the MPA-JHU catalogue for the JTM12 data, and from mass-to-light ratios obtained using the \citet{B03} (g-r)-$M_{*}/L_{g}$ relation for the \citet{GFS09} data. The additional observational data (green points), kindly provided by C. Conroy \& G. Graves (priv. comm.), are drawn from the SDSS-DR7, selecting galaxies in the redshift range $0.025<z<0.06$ with bulge-to-total light ratios $\geq 0.7$ and (g-r) $\geq 0.051\:\textnormal{log}(M_{*}) + 0.14$. These galaxies are binned by $M_{*}$ and their stacked spectra are used to obtain element enhancements using the SPS models of \citet{CvD12a,CGvD13}. For a detailed comparison of these different methods, see \S 6 of \citet{CGvD13}.

Looking at the panels in Fig. \ref{fig:Ellip_XFe_Mstar} individually, we can see that the slopes of the model relations for [C/Fe] and [N/Fe] are shallower than observed. This is discussed further in \S \ref{sec:Carbon and Nitrogen}. Although our PL set-up reproduces a slope of the $M_{*}$-[O/Fe] relation for the mass-age-selected sub-sample close to that obtained by JTM12 (as also shown in Fig. \ref{fig:Ellip_OFe_Mstar_allDTDs}), the newly-selected SDSS-DR7 data suggests a significantly steeper slope. Steeper slopes are obtained in the model when either using the Gaussian (delayed only) DTD, or when allowing direct ejection of light $\alpha$ elements out of galaxies via galactic winds, as explained in \S \ref{sec:Galactic winds}. An increase in $\tau_{\textnormal{min}}$ (the start time for SN-Ia explosions, see \S \ref{sec:SN-Ia DTD}) also slightly increases the slope. For example, increasing $\tau_{\textnormal{min}}$ from 35 to 45 Myrs increases the slope of the $M_{*}$-[O/Fe] relation by $\sim 0.004$ when using the power-law DTD.

Our PL set-up produces a flat $M_{*}$-[Mg/Fe] relation, with a slope slightly shallower than the well-constrained relation obtained from the SDSS-DR7 data. The SDSS-DR7 $M_{*}$-[Mg/Fe] relation, in turn, is flatter than the other observational data sets we compare to here. In our model, the slope of the $M_{*}$-[Mg/Fe] relation is flatter than the other light $\alpha$ elements because Mg is produced in greater amounts by low-metallicity SNe-II than high-metallicity SNe-II when using the P98 yields (compare the bottom two panels in Fig. \ref{fig:P98 SN-II}). This is not the case for the CL04 SN-II yields, which produce a slope for [Mg/Fe] very similar to that of [O/Fe], due to the negligable difference in their metallicity dependence.

Strong, positive slopes for the heavier $\alpha$ elements (Si, S and Ca) are obtained for \textit{all} three of the DTDs considered here when using the P98 SN-II yields. This is because these elements are produced only in small amounts by SSPs at low metallicity, as they are easily locked into the stellar remnants of the most massive, low-metallicity SNe-II (see \S \ref{sec:SN-II yields}). Low-mass elliptical galaxies, which never obtain high stellar metallicities, therefore never produce enough Si, S or Ca to obtain high enhancements. As the CL04 SN-II yields do not take account of the prior stellar winds' effect on remnant composition, the slopes produced for the heavier $\alpha$ elements are very similar to those of the lighter $\alpha$ elements. This means that \textit{all} slopes are equally sensitive to the choice of DTD when using the CL04 yields, such that positive slopes are \textit{only} obtained if a minor prompt component of SNe-Ia is assumed. We note that the slightly shallower slopes for $M_{*}$-[Ca/Fe] observed in the real Universe may indicate that a larger proportion of heavy $\alpha$ elements come from SNe-Ia than is the case in our model (see \citealt{CGvD13}).

\subsubsection{Galactic winds} \label{sec:Galactic winds}
\begin{figure}
\centering
\includegraphics[totalheight=0.2\textheight, width=0.46\textwidth]{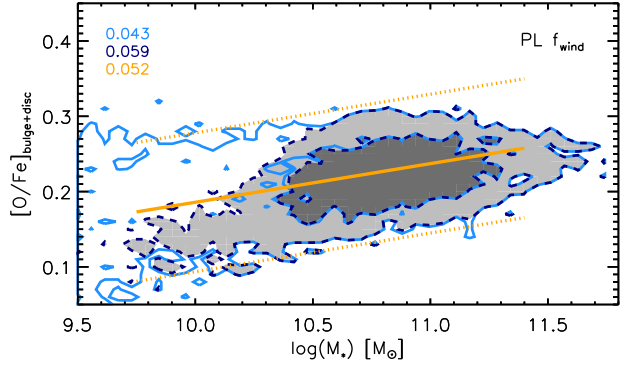} \\
\caption{The $M_{*}$-[O/Fe] relation for our model elliptical sample (when using a power-law SN-Ia DTD), with the SN feedback scheme that allows some direct ejection of light $\alpha$ elements out of galaxies via galactic winds (see \S \ref{sec:Galactic winds}). Contours and lines are as in Fig. \ref{fig:Ellip_OFe_Mstar_allDTDs}.}
\label{fig:OFe_fwind}
\end{figure}

The slopes of the [$\alpha$/Fe] relations, for \textit{all} SN-Ia DTDs and SN-II yields considered, are strengthened by introducing metal-rich winds, which suppress the enhancements in low-mass ellipticals. Currently, \textsc{L-Galaxies} does not invoke direct ejection of material by SNe, instead always fully mixing SN ejecta with the galaxy's ISM before reheating a fraction of this enriched gas into the CGM. However, a simple wind model, which allows a fraction of the material and energy ejected by SN-II explosions in the disc to be deposited directly into the hot gas, increases the slopes of the $M_{*}$-[$\alpha$/Fe] relations. This is shown for our PL set-up in Fig. \ref{fig:OFe_fwind}. This figure can be compared to the middle panel of Fig. \ref{fig:Ellip_OFe_Mstar_allDTDs}.

A scheme where only SNe-II are expected to form a collimated galactic wind is physically motivated by the fact that metal-rich winds (ubiquitous in local, star-forming galaxies) appear to be oxygen rich, $\alpha$ enhanced, and occur shortly after bouts of star formation (e.g. \citealt{MKH02,T11}). This scheme also allows the remainder of the mass and energy returned by disc SNe-II to mix with and reheat cold gas.

\begin{figure}
\centering
\includegraphics[totalheight=0.25\textheight, width=0.31\textwidth]{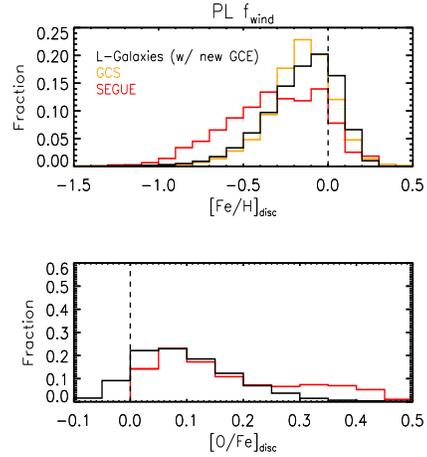}
\caption{The [Fe/H] and [O/Fe] distributions for model MW-type galaxies (PL set-up), with the alternative wind scheme (see \S \ref{sec:Galactic winds}). This can be compared to the middle panel of Fig. \ref{fig:MW_FeH_and_OFe_dists_FP_ObsComp}). Although the [Fe/H] distribution is not significantly affected when including galactic winds, there is a slight increase in the number of low-[O/Fe] SSPs formed.}
\label{fig:FeH_OFe_wObs_fwind}
\end{figure}

We set the fraction of material from disc SN-II that is directly ejected via the wind to be inversely proportional to the cold gas surface density of the ISM through which it must pass;

\begin{equation} \label{eqn:fwind}
f_{\textnormal{wind}} = \textnormal{min}\left[1.0, \left(\frac{\Sigma_{\textnormal{cold}}}{10\;\textnormal{M}_{\textnormal{\astrosun}} \textnormal{pc}^{-2}}\right)^{-1}\right] \;\;.
\end{equation}
A similar dependency on the gas surface density has also been used in the smoothed-particle hydrodynamical simulations of \citet{HQM12} and in the \textsc{Galform} semi-analytic model by \citet{LLB13} (but see \citealt{N12}). Interestingly, our preferred characteristic gas surface density of $\sim10$ $\textnormal{M}_{\textnormal{\astrosun}} \textnormal{pc}^{-2}$, below which all SN-II ejecta material is put into the wind, is very close to that below which the SFR surface density drops in local, spiral galaxies (e.g. \citealt{B08,BLW11}).

This simple wind scheme lowers the stellar $\alpha$ enhancements of low-mass ellipticals more than high-mass ellipticals, because low-mass ellipticals tend to have lower-density ISM, and so can dump their SN-II ejecta more efficiently into the CGM. This does \textit{not} significantly affect the [Fe/H] distribution in the discs of MW-type model galaxies, although the number of low-[O/Fe] SSPs does increase slightly, as shown in Fig. \ref{fig:FeH_OFe_wObs_fwind} (compare to the middle panel of Fig. \ref{fig:MW_FeH_and_OFe_dists_FP_ObsComp}). However, this simple wind scheme \textit{does} cause a significant under-enrichment of the ISM in low-mass star-forming galaxies by $z=0$, which steepens the slope of the model $M_{*}$-$Z_{\textnormal{cold}}$ relation away from that seen in observations. Therefore, although we show here that metal-rich winds are a way of strengthening positive slopes in the $M_{*}$-[$\alpha$/Fe] relations of ellipticals, we do \textit{not} claim that our simple wind model can solve all the problems of GCE modelling.

\subsubsection{Carbon and Nitrogen} \label{sec:Carbon and Nitrogen}
The case of C and N is more complicated than that of other heavy elements, not least because N is both a primary and secondary element. Our model produces slopes for C and N that are flatter than for the $\alpha$ elements (see top two panels in Fig. \ref{fig:Ellip_XFe_Mstar}). This is to be expected if C and N are predominantly released in AGB winds (as they are at low metallicity in our model), yet observations suggest that these enhancements should also produce positive slopes. To further compound the problem, observations by JTM12 of the $M_{*}$-[C/O] and $M_{*}$-[N/O] relations show that these also have positive slopes.

\begin{figure}
\centering
\includegraphics[totalheight=0.2\textheight, width=0.46\textwidth]{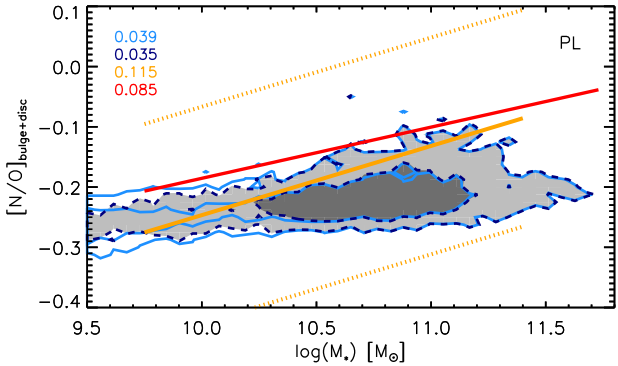} \\
\caption{The $M_{*}$-[N/O] relation for our model elliptical sample (PL set-up). Contours and lines are as in Fig. \ref{fig:Ellip_OFe_Mstar_allDTDs}, plus a fit to the model relation when increasing the yield of nitrogen from high-metallicity SNe-II by a factor of 1.5 (see \S \ref{sec:Carbon and Nitrogen}) (red line). We find that a simple (although \textit{ad hoc}) increase in the nitrogen yield is enough to obtain a strong, positive slope in this relation.}
\label{fig:M-NO_Ellips_PowerLaw_P98_nofwind_A005_2e51_3-16_mk2}
\end{figure}

Fig. \ref{fig:M-NO_Ellips_PowerLaw_P98_nofwind_A005_2e51_3-16_mk2} shows the $M_{*}$-[N/O] relation for our PL model (without additional galactic winds), along with the fit to observations from JTM12. This model relation is also flatter than observed, and its slope \textit{increases} with the amount of prompt SNe-Ia in the DTD. We note that there is a scatter of high-mass model galaxies with [N/O] values more similar to those observed. However, these higher-[N/O] galaxies tend to be young for their mass in the model, whereas the opposite is true in the observational sample.

One way to increase the slopes in both the $M_{*}$-[N/Fe] and $M_{*}$-[N/O] relations is to assume a greater amount of N production in high-metallicity \textit{massive} stars, as suggested by JTM12. Doing so implies a boost in secondary nitrogen production. Given the current uncertainty in the amount of secondary N production in stars, this \textit{could} be a plausible solution, although this is far from certain. The red line in Fig. \ref{fig:M-NO_Ellips_PowerLaw_P98_nofwind_A005_2e51_3-16_mk2} is a fit to the full model elliptical sample when arbitrarily increasing the N released by SNe-II of metallicity $\geq 0.02$ by a factor of 1.5. A similar increase is also seen in the slope of the $M_{*}$-[N/Fe] relation. Although this is an \textit{ad hoc} adjustment made to the stellar yields, it does indicate that such a change is capable of improving the values of both [N/Fe] and [N/O] in our model ellipticals.

When using the CL04 yields, the $M_{*}$-[C/Fe] and $M_{*}$-[N/Fe] relations are even flatter and the $M_{*}$-[N/O] relation has a negative slope, due to the lower production of C and N by SNe-II that these yield tables infer. This, again, is in contradiction with the observational data considered, suggesting that the P98 yields, which take account of prior stellar mass loss from massive stars (and so are more dependent on initial mass and metallicity), produce more realistic results in our GCE model.

\textit{} \newline
To conclude this section, we reiterate that positive slopes in the $M_{*}$-[$\alpha$/Fe] relations of local elliptical galaxies can be obtained if either a SN-Ia DTD with minor prompt component (i.e. $\leq 50$ per cent within $\sim 400$ Myrs) is used, and/or galactic winds driven by SNe-II are allowed to directly eject metals out of galaxies.

\section{Conclusions} \label{sec:Conclusions}
We have implemented a new GCE model into the Munich semi-analytic model, \textsc{L-Galaxies}, which accounts for the delayed enrichment of a series of heavy elements from SNe-II, SNe-Ia and AGB stars. We have also compared the results of this implementation with the chemical composition of local, star-forming galaxies, the MW stellar disc, and local, elliptical galaxies. Our conclusions are as follows:

\begin{itemize}
\item The gas-phase mass-metallicity relation for local, star-forming galaxies (when using the Bayesian, SDSS metallicities of \citealt{T04}) is very well reproduced by our new model. However, we caution that both the slope and amplitude of the observed $M_{*}$-$Z_{\textnormal{g}}$ relation depend strongly of the metallicity diagnostic chosen (e.g. \citealt{KE08}). The stellar components of low-mass, star-forming galaxies tend to be more metal-rich in our model than observed.
\item The [Fe/H] distribution of G dwarfs in the MW disc is reasonably reproduced by our model, for all forms of SN-Ia DTD we consider. However, the high-[O/Fe] tail in the MW [O/Fe] distribution is best reproduced when using a SN-Ia DTD with a \textit{minor} prompt component (i.e. $\leq 50$ per cent within $\sim 400$ Myrs), such as a Gaussian DTD centered on $\sim 1$ Gyr, or a power-law DTD with slope $\sim -1.12$ and $\tau_{\textnormal{min}}\gtrsim 35$ Myrs.

\item Positive slopes in the $M_{*}$-[$\alpha$/Fe] relations of local, elliptical galaxies are also obtained when assuming a minor prompt component (i.e. $\leq 50$ per cent within $\sim 400$ Myrs). The strength of the slope is inversely proportional to the fraction of prompt SNe-Ia. These results are achieved when using the same implementation which produces our star-forming-galaxy and MW results.

\item The inclusion of metal-rich galactic winds, driven by SN-II explosions, strengthens the positive slopes in the $M_{*}$-[$\alpha$/Fe] relations of ellipticals for \textit{all} forms of SN-Ia DTD and SN-II yields considered. However, our simple, ISM-density-dependent wind scheme reduces the gas-phase metallicity of low-mass, star-forming galaxies, and so does not fully solve the problem.

\item There is a clear correlation between mass, age and $\alpha$ enhancement in our model. This, along with the above findings, suggests that the chemical compositions of a diverse array of galaxies can be reconciled without requiring a variable IMF. Although an IMF that varied with SFR would likely produce similar results, it is instructive to see that this is not the only solution, given that the true behaviour of the stellar IMF is still uncertain.

\item Overall, our results suggest, given the assumptions and limitations discussed, that the best model for matching the wide range of observational data considered here should include a) a power-law SN-Ia DTD, b) SN-II yields that take account of prior mass loss through stellar winds, and c) some direct ejection of light $\alpha$ elements into the CGM.
\end{itemize}

We conclude by highlighting two unavoidable limitations of this work and GCE models in general. First, the stellar yields used as an input have a strong influence on the results, as shown here by comparing the SN-II yields of P98 and CL04, and also in other studies (e.g. \citealt{F04,R10}). Until the true yields ejected by stars of all masses and metallicities are better understood, the accuracy of GCE models will always be uncertain. Second, we only consider one free GCE parameter in this work, $A$, the fraction of objects in a stellar SSP in the range $3\leq M/\textnormal{M}_{\textnormal{\astrosun}}\leq 16$ that are SN-Ia progenitors. However, the values assumed for other GCE parameters (such as $\tau_{\textnormal{min}}$, $M_{\textnormal{max}}$, or the high-mass slope of the IMF) could also take on different values in reality. If so, our tuning of $A$ could also be correcting for these other uncertainties, and would not solely represent the efficiency of SN-Ia-progenitor formation. Further testing against additional observational data from Local Group dwarf galaxies, the intracluster medium of galaxy clusters, and outliers from the mass-metallicity relation, should help better constrain such uncertainties. These topics will be the focus of future work.

\section*{Acknowledgments} \label{sec:Acknowledgements}
The authors would like to thank Jo Bovy, Francesco Calura, Charlie Conroy, Claudio Dalla Vecchia, Anna Gallazzi, Genevieve Graves, Chervin Laporte, Pierre Maggi, Hans-Walter Rix, Ashley Ruiter, Ivo Seitenzahl, Christy Tremonti, Lizhi Xie, Rob Wiersma and Tyrone Woods for invaluable discussions during the undertaking of this work. RMY acknowledges the financial support of the Deutsche Forschungsgesellshaft (DFG). PAT acknowledges support from the Science and Technology Facilities Council (grant number ST/I000976/1). SW was supported in part by Advanced Grant 246797 ``GALFORMOD'' from the European Research Council.


\begin{thebibliography}{99}
\bibitem[\protect\citeauthoryear{Aller}{1942}]{A42} Aller L. H., 1942, ApJ, 95, 52
\bibitem[\protect\citeauthoryear{Anders \& Grevesse}{1989}]{AG89} Anders E, Grevesse N. 1989. Geochim. Cosmochim. Acta 53:197–214
\bibitem[\protect\citeauthoryear{Arrigoni et al.}{2010a}]{A10a} Arrigoni M., Trager S. C., Somerville R. S., Gibson B. K., 2010a, MNRAS, 402, 173
\bibitem[\protect\citeauthoryear{Arrigoni et al.}{2010b}]{A10b} Arrigoni M., Trager S. C., Somerville R. S., 2010b, arXiv:1006.1147 [astro-ph]
\bibitem[\protect\citeauthoryear{Asplund et al.}{2009}]{A09} Asplund M., Grevesse N., Sauval A. J., Scott P., 2009, ARA\&A, 47, 481
\bibitem[\protect\citeauthoryear{Bastian, Covey \& Meyer}{2010}]{BCM10} Bastian N., Covey K. R., Meyer M. R., 2010, ARA\&A, 48, 339
\bibitem[\protect\citeauthoryear{Bell et al.}{2003}]{B03} Bell E. F., McIntosh D. H., Katz N., Weinberg M. D., 2003, ApJS, 149, 289
\bibitem[\protect\citeauthoryear{Bigiel et al.}{2008}]{B08} Begiel F., Leroy A., Walter F., Brinks E., de Blok W. J. G., Madore B., Thornley M. D., 2008, AJ, 136, 2846
\bibitem[\protect\citeauthoryear{Bigiel, Leroy \& Walter}{2011}]{BLW11} Begiel F., Leroy A., Walter F., 2010, IAUS, 270, 327B
\bibitem[\protect\citeauthoryear{Bovy et al.}{2012a}]{B12a} Bovy J., Rix H.-W., Hogg D. W., 2012a, ApJ, 751, 131
\bibitem[\protect\citeauthoryear{Bovy et al.}{2012b}]{B12b} Bovy J., Rix H.-W., Liu C., Hogg D. W., Beers T. C., Lee Y. S., 2012b, ApJ, 753, 148
\bibitem[\protect\citeauthoryear{Brandt et al.}{2010}]{B10} Brandt T. D., Tojeiro R., Aubourg \'{E}, Heavens A., Jimenez R., Strauss M. A. , 2010, AJ, 140, 804
\bibitem[\protect\citeauthoryear{Bruzual \& Charlot}{2003}]{BC03} Bruzual A. G., Charlot S., 2003, MNRAS, 344, 1000	
\bibitem[\protect\citeauthoryear{Burbidge et al.}{1957}]{BBFH57} Burbidge E. M., Burbidge G. R., Fowler W. A., \& Hoyle F., 1957, Reviews of Modern Physics,
29, 547
\bibitem[\protect\citeauthoryear{Calura et al.}{2010}]{C10} Calura F., Recchi S., Matteucci F., Kroupa P., 2010, MNRAS, 406, 1985
\bibitem[\protect\citeauthoryear{Calura et al.}{2012}]{C12} Calura F., et al., 2012, MNRAS, 427, 1401
\bibitem[\protect\citeauthoryear{Calura \& Menci}{2009}]{CM09} Calura F., Menci N., 2009, MNRAS, 400, 1347
\bibitem[\protect\citeauthoryear{Calura \& Menci}{2011}]{CM11} Calura F., Menci N., 2011, MNRAS, 413, 1
\bibitem[\protect\citeauthoryear{Chamberlain \& Aller}{1951}]{CA51} Chamberlain J. W., Aller L. H., 1951, ApJ, 114, 52
\bibitem[\protect\citeauthoryear{Chabrier}{2003}]{C03} Chabrier G., 2003, PASP, 115, 763
\bibitem[\protect\citeauthoryear{Chieffi \& Limongi}{2004}]{CL04} Chieffi A., Limongi M., 2004, ApJ, 608, 405, (CL04)
\bibitem[\protect\citeauthoryear{Conroy \& van Dokkum}{2012a}]{CvD12a} Conroy C., van Dokkum P. G., 2012a, ApJ, 747, 69
\bibitem[\protect\citeauthoryear{Conroy \& van Dokkum}{2012b}]{CvD12b} Conroy C., van Dokkum P. G., 2012b, ApJ, 760, 71
\bibitem[\protect\citeauthoryear{Conroy, Graves \& van Dokkum}{2013}]{CGvD13} Conroy C., Graves G. J., van Dokkum P. G., 2013, arXiv:1303.6629 [astro-ph]
\bibitem[\protect\citeauthoryear{Croton et al.}{2006}]{C06} Croton D. J., et al., 2006, MNRAS, 365, 11
\bibitem[\protect\citeauthoryear{Dellenbusch, Gallagher \& Knezek}{2007}]{DGK07} Dellenbusch K. E., Gallagher III, J. S., Knezek P. M., 2007, ApJ, 655, L29
\bibitem[\protect\citeauthoryear{De Lucia, Kauffmann \& White}{2004}]{DL04} De Lucia G., Kauffmann G., White S. D. M., 2004, MNRAS, 349, 1101
\bibitem[\protect\citeauthoryear{De Lucia et al.}{2006}]{DL06} De Lucia G., Springel V., White S. D. M., Croton D., Kauffmann G., 2006, MNRAS, 366, 499
\bibitem[\protect\citeauthoryear{De Lucia \& Blaizot}{2007}]{DLB07} De Lucia G., Blaizot J., 2007, MNRAS, 375, 2
\bibitem[\protect\citeauthoryear{de Plaa et al.}{2007}]{dP07} de Plaa J., Werner N., Bleeker J. A. M., Vink J., Kaastra J. S., M\'{e}ndez M., 2007, A\&A, 465, 345
\bibitem[\protect\citeauthoryear{De Rossi et al.}{2009}]{DR09} De Rossi M. E., Tissera P. B., De Lucia G., Kauffmann G., 2009, MNRAS, 395, 210
\bibitem[\protect\citeauthoryear{Eddington}{1920}]{E20} Eddington A. S., 1920, Observatory, 43, 341
\bibitem[\protect\citeauthoryear{Edvardsson et al.}{1993}]{E93} Edvardsson B., Andersen J., Gustafsson B., Lambert D. L., Nissen P. E., Tomkin J., 1993, A\&A, 275, 101
\bibitem[\protect\citeauthoryear{Elmegreen}{2006}]{El06} Elmegreen B. G., 2006, ApJ, 648, 572
\bibitem[\protect\citeauthoryear{Fran\c{c}ois et al.}{2004}]{F04} Fran\c{c}ois P., Matteucci F., Cayrel R., Spite M., Spite F., Chiappini C., 2004, A\&A, 421, 613
\bibitem[\protect\citeauthoryear{Friel}{1995}]{F95} Friel E. D., 1995, ARA\&A, 33, 381
\bibitem[\protect\citeauthoryear{Fu et al.}{2013}]{F13} Fu J., Kauffmann G., Huang M., Yates R. M., Moran S., Heckman T. M., Dav\'{e} R., Guo Q., 2013, arXiv:1303.5586 [astro-ph]
\bibitem[\protect\citeauthoryear{Fumagalli, Da Silva \& Krumholz}{2011}]{FDK11} Fumagalli M., da Silva R. L., Krumholz M. R., 2011, ApJ, 741L, 26
\bibitem[\protect\citeauthoryear{Gallazzi et al.}{2005}]{G05} Gallazzi A., Charlot S., Brinchmann J., White S. D. M., Tremonti C. A., 2005, MNRAS, 362, 41
\bibitem[\protect\citeauthoryear{Gnedin}{2000}]{G00} Gnedin N. Y., 2000, ApJ, 542, 535
\bibitem[\protect\citeauthoryear{Graves, Faber \& Schiavon}{2009}]{GFS09} Graves G. J., Faber S. M., Schiavon R. P., 2009, ApJ, 698, 1590
\bibitem[\protect\citeauthoryear{Grevesse, Noels \& Sauval}{1996}]{GNS96} Grevesse N., Noels A., Sauval A. J., 1996, ASP Conference Series, Volume 99, 117
\bibitem[\protect\citeauthoryear{Greggio \& Renzini}{1983}]{GR83} Greggio L., Renzini A., 1983, A\&A, 118, 217
\bibitem[\protect\citeauthoryear{Greggio}{2005}]{Gr05} Greggio L., 2005, A\&A, 441, 1055
\bibitem[\protect\citeauthoryear{Gunawardhana et al.}{2011}]{G11} Gunawardhana M. L. P., et al., 2011, MNRAS, 415, 1647
\bibitem[\protect\citeauthoryear{Guo et al.}{2011}]{G10} Guo Q., White S. D. M., Boylan-Kolchin M. De Lucia G., Kauffmann G., Lemson G., Li C., Springel V., Weinmann S., 2011, MNRAS, 413, 101
\bibitem[\protect\citeauthoryear{Guo et al.}{2013}]{G13} Guo Q., White S. D. M., Angulo R. E., Henriques B., Lemson G., Boylan-Kolchin M., Thomas P., Short C., 2013, MNRAS, 428, 1351
\bibitem[\protect\citeauthoryear{Henriques et al.}{2013}]{H13} Henriques B., White S. D. M., Thomas P. A., Angulo R. E., Guo Q., Lemson G., Springel V., 2013, arXiv:1212.1717 [astro-ph]
\bibitem[\protect\citeauthoryear{Henriques \& Thomas}{2010}]{HT10} Henriques B. M. B., Thomas P. A., 2010, MNRAS, 403, 768
\bibitem[\protect\citeauthoryear{Holmberg, Nordstr\"{o}m \& Andersen}{2009}]{HNA09} Holmberg J., Nordstr\"{o}m B., Andersen J., 2009, A\&A, 501, 941
\bibitem[\protect\citeauthoryear{Hopkins, Quataert \& Murray}{2012}]{HQM12} Hopkins P. F., Quataert E., Murray N., 2012, MNRAS, 421, 3522
\bibitem[\protect\citeauthoryear{Iben \& Tutukov}{1984}]{IT84} Iben I. Jr., Tutukov A. V., 1984, ApJS, 54, 335
\bibitem[\protect\citeauthoryear{Johansson, Thomas \& Maraston}{2012}]{JTM12} Johansson J., Thomas D., Maraston C., 2012, MNRAS, 421, 1908, (JTM12)
\bibitem[\protect\citeauthoryear{Kauffmann, White \& Guideroni}{1993}]{KWG93} Kauffmann G., White S. D. M., Guideroni B., 1993, MNRAS, 264, 201
\bibitem[\protect\citeauthoryear{Kauffmann et al.}{1999}]{K99} Kauffmann G., Colberg J. M., Diaferio A., White S. D. M., 1999, MNRAS, 307, 529
\bibitem[\protect\citeauthoryear{Kewley \& Ellison}{2008}]{KE08} Kewley L. J., Ellison S. L., 2008, ApJ, 681, 1183
\bibitem[\protect\citeauthoryear{Kroupa}{2001}]{K01} Kroupa P., 2001, MNRAS, 322, 231
\bibitem[\protect\citeauthoryear{Lagos, Lacey \& Baugh}{2013}]{LLB13} Lagos C. d. P., Lacey C. G., C. M. Baugh, 2013, arXiv:1303.6635 [astro-ph]
\bibitem[\protect\citeauthoryear{Lequeux et al.}{1979}]{L79} Lequeux J., Peimbert M., Rayo J. F., Serrano A., Torres-Peimbert S., 1979, A\&A, 80, 155L
\bibitem[\protect\citeauthoryear{Lia, Portinari \& Carraro}{2002}]{LPC02} Lia C., Portinari L., Carraro G., 2002, MNRAS, 330, 821
\bibitem[\protect\citeauthoryear{Maeder}{1992}]{M92} Maeder A., 1992, A\&A, 264, 105
\bibitem[\protect\citeauthoryear{Mannucci, Della Valle \& Panagia}{2006}]{MVP06} Mannucci F., Della Valle M., Panagia N., 2006, MNRAS, 370, 773
\bibitem[\protect\citeauthoryear{Matteucci}{1986}]{M86} Matteucci F., 1986, MNRAS, 221, 911
\bibitem[\protect\citeauthoryear{Matteucci \& Greggio}{1986}]{MG86} Matteucci F., Greggio L., 1986, A\&A, 154, 279
\bibitem[\protect\citeauthoryear{Matteucci}{1994}]{M94} Matteucci F., 1994, A\&A, 288, 57
\bibitem[\protect\citeauthoryear{Matteucci \& Recchi}{2001}]{MR01} Matteucci F., Recchi S., 2001, ApJ, 558, 351
\bibitem[\protect\citeauthoryear{Matteucci et al.}{2006}]{M06} Matteucci F., Panagia N., Pipino A., Mannucci F., Recchi S., Della Valle M., 2006, MNRAS, 372, 265
\bibitem[\protect\citeauthoryear{Matteucci et al.}{2009}]{M09} Matteucci F., Spitoni E., Recchi S., Valiante R., 2009, A\&A, 501, 531
\bibitem[\protect\citeauthoryear{Maoz \& Badenes}{2010}]{MB10} Maoz D., Badenes C., 2010, MNRAS, 407, 1314 %arXiv:1003.3031 [astro-ph]
\bibitem[\protect\citeauthoryear{Maoz Sharon \& Gal-Yam}{2010}]{MSG10} Maoz D., Sharon K, Gal-Yam A., 2010, ApJ, 722, 1879
\bibitem[\protect\citeauthoryear{Maoz \& Mannucci}{2012}]{MM12} Maoz D., Mannucci F., 2012, PASA, 29, 447 %arXiv:1111.4492 [astro-ph]
\bibitem[\protect\citeauthoryear{Maoz, Mannucci \& Brandt}{2012}]{MMB12} Maoz D., Mannucci F., Brandt T. D., 2012, MNRAS, 426, 3282 %arXiv:1206.0465 [astro-ph]
\bibitem[\protect\citeauthoryear{Marigo}{2001}]{M01} Marigo P., 2001, A\&A, 370, 194, (M01)
\bibitem[\protect\citeauthoryear{Maraston}{2005}]{M05} Maraston C., 2005, MNRAS, 362, 799
\bibitem[\protect\citeauthoryear{Martin, Kobulnicky \& Heckman}{2002}]{MKH02} Martin C. L., Kobulnicky H. A., Heckman T. M., 2002, ApJ, 574, 663
\bibitem[\protect\citeauthoryear{McWilliam}{1997}]{McW97} McWilliam, ARA\&A, 35, 503
\bibitem[\protect\citeauthoryear{Nagashima et al.}{2005b}]{N05b} Nagashima M., Lacey C. G., Okamoto T., Baugh C. M., Frenk C. S., Cole S., 2005, MNRAS, 363, L31
\bibitem[\protect\citeauthoryear{Newman et al.}{2012}]{N12} Newman S. F., et al., 2012, ApJ, 761, 43
\bibitem[\protect\citeauthoryear{Nomoto et al.}{1984}]{N84} Nomoto K., Thielemann F.-K., Yokoi K., 1984, ApJ, 286, 644
\bibitem[\protect\citeauthoryear{Nordstr\"{o}m et al.}{2004}]{N04} Nordstr\"{o}m B., et al., 2004, A\&A, 418, 989
\bibitem[\protect\citeauthoryear{Okamoto, Gao \& Theuns}{2008}]{OGT08} Okamoto T., Gao L., Theuns T., 2008, MNRAS, 390, 920
\bibitem[\protect\citeauthoryear{Padovani \& Matteucci}{1993}]{PM93} Padovani P., Matteucci M., 1993, ApJ, 416, 26
\bibitem[\protect\citeauthoryear{Panter et al.}{2008}]{P08} Panter B., Jimenez R., Heavens A. F., Charlot S., 2008, MNRAS, 391, 1117
\bibitem[\protect\citeauthoryear{Peeples, Pogge \& Stanek}{2008}]{PPS08} Peeples M. S., Pogge R. W., Stanek K. Z., 2008, ApJ, 685, 904
\bibitem[\protect\citeauthoryear{Pilkington et al.}{2012}]{P12} Pilkington K., et al., 2012, MNRAS, 425, 969
\bibitem[\protect\citeauthoryear{Pipino et al.}{2009a}]{P09a} Pipino A., Chiappini C., Graves G., Matteucci F., 2009a, MNRAS, 396, 1151
\bibitem[\protect\citeauthoryear{Pipino et al.}{2009b}]{P09b} Pipino A., Devriendt J. E. G., Thomas D., Silk J., Kaviraj S., 2009b, A\&A, 505, 1075
\bibitem[\protect\citeauthoryear{Pipino \& Matteucci}{2004}]{PM04} Pipino A., Matteucci F., 2004, MNRAS, 347, 968
\bibitem[\protect\citeauthoryear{Pipino \& Matteucci}{2011}]{PM11} Pipino A., Matteucci F., 2011, A\&A, 530, 98
\bibitem[\protect\citeauthoryear{Portinari, Chiosi \& Bressan}{1998}]{P98} Portinari L., Chiosi C., Bressan A., 1998, A\&A, 334, 505, (P98)
\bibitem[\protect\citeauthoryear{Romano et al.}{2010}]{R10} Romano D., Karakas A. I., Tosi M., Matteucci F., 2010, A\&A, 522, 32
\bibitem[\protect\citeauthoryear{Ruiter et al.}{2011}]{R11} Ruiter A. J., Belczynski K., Sim S. A., Hillebrandt W., Fryer C. L., Fink M., Kromer M., 2011, MNRAS, 417, 408
\bibitem[\protect\citeauthoryear{Salaris \& Cassisi}{2005}]{SC05} Salaris M., Cassisi S., 2005, 'Evolution of Stars and Stellar Populations', Wiley, Chichester
\bibitem[\protect\citeauthoryear{Salpeter}{1955}]{S55} Salpeter E. E., 1955, ApJ, 121, 161
\bibitem[\protect\citeauthoryear{Scalo}{1986}]{S86} Scalo J. M., 1986, Fundamentals of Cosmic Physics, Volume 11, 1
\bibitem[\protect\citeauthoryear{Schiavon}{2007}]{S07} Schiavon R. P., 2007, ApJS, 171, 146
\bibitem[\protect\citeauthoryear{Seitenzahl et al.}{2013}]{S13} Seitenzahl I. R., et al., 2013, MNRAS, 429, 1156
\bibitem[\protect\citeauthoryear{Serra \& Trager}{2007}]{ST07} Serra P., Trager S. C., 2007, MNRAS, 374, 769
\bibitem[\protect\citeauthoryear{Spergel et al.}{2003}]{S03} Spergel D. N., et al., 2003, ApJS, 148, 175
\bibitem[\protect\citeauthoryear{Springel et al.}{2001}]{S01} Springel V., White S. D. M., Tormen G., Kauffmann G., 2001, MNRAS, 328, 726
\bibitem[\protect\citeauthoryear{Springel et al.}{2005}]{S05} Springel V., et al., 2005, Nature, 435, 629
\bibitem[\protect\citeauthoryear{Strolger et al.}{2004}]{S04} Strolger L.-G., et al., 2010, ApJ, 613, 200
\bibitem[\protect\citeauthoryear{Sutherland \& Dopita}{1993}]{SD93} Sutherland R. S., Dopita M. A., 1993, ApJS, 88, 253
\bibitem[\protect\citeauthoryear{Thielemann et al.}{2003}]{T03} Thielemann F.-K., et al., 2003, `From Twilight to Highlight: The Physics of Supernov\ae{}, Supernova Nucleosynthesis and Galactic Evolution.'
\bibitem[\protect\citeauthoryear{Thomas, Greggio \& Bender}{1998}]{TGB98} Thomas D., Greggio L., Bender R., 1998, MNRAS, 296, 119
\bibitem[\protect\citeauthoryear{Thomas \& Kauffmann}{1999}]{TK99} Thomas D., Kauffmann G., 1999, ASPC, 192, 261
\bibitem[\protect\citeauthoryear{Thomas, Greggio \& Bender}{1999}]{TGB99} Thomas D., Greggio L., Bender R., 1999, MNRAS, 302, 537
\bibitem[\protect\citeauthoryear{Thomas}{1999}]{T99} Thomas D., 1999, MNRAS, 306, 655
\bibitem[\protect\citeauthoryear{Thomas et al.}{2010}]{T10} Thomas D., Maraston C., Schawinski K, Sarzi M., Silk J., 2010, MNRAS, 404, 1775
\bibitem[\protect\citeauthoryear{Thomas, Maraston \& Johansson}{2011b}]{TMJ11b} Thomas D., Maraston C., Johansson J., 2011b, MNRAS, 412, 2183
\bibitem[\protect\citeauthoryear{Tinsley}{1980}]{T80} Tinsley B. M., 1980, Fundamentals of Cosmic Physics, Volume 5, 287
\bibitem[\protect\citeauthoryear{Tissera, White \& Scannapieco}{2012}]{TWS12} Tissera P. B., White S. D. M., Scannapieco C., 2012, MNRAS, 420, 255
\bibitem[\protect\citeauthoryear{Totani et al.}{2008}]{T08} Totani T., Morokuma T., Oda T., Doi M., Yasuda N., 2008, PASJ, 60, 1327
\bibitem[\protect\citeauthoryear{Tremonti et al.}{2004}]{T04} Tremonti C. A., et al., 2004, ApJ, 613, 898
\bibitem[\protect\citeauthoryear{Tumlinson et al.}{2011}]{T11} Tumlinson J., et al., 2011, Science, 334, 948
\bibitem[\protect\citeauthoryear{van Dokkum \& Conroy}{2010}]{vDC10} van Dokkum P. G., Conroy C., 2010, Nature, 468, 940
\bibitem[\protect\citeauthoryear{Walch et al.}{2011}]{W11} Walch S., W\"{u}nsch R., Burkert A., Glover S., Whitworth A., 2011, ApJ, 733, 47
\bibitem[\protect\citeauthoryear{Webbink}{1984}]{W84} Webbink R. F., 1984, ApJ, 277, 355
\bibitem[\protect\citeauthoryear{Weidner \& Kroupa}{2006}]{WK06} Weidner C., Kroupa P., 2006, MNRAS, 365, 1333
\bibitem[\protect\citeauthoryear{Weinmann et al.}{2006}]{W06} Weinmann S. M., van den Bosch F. C., Yang X., Mo H. J., Croton D. J., Moore B., 2006, MNRAS, 372, 1161
\bibitem[\protect\citeauthoryear{Whelan \& Iben}{1973}]{WI73} Whelan J., Iben I. Jr., 1973, ApJ, 186, 1007
\bibitem[\protect\citeauthoryear{White \& Frenk}{1991}]{WF91} White S. D. M., Frenk C., 1991, ApJ, 379, 52
\bibitem[\protect\citeauthoryear{Wiersma et al.}{2009b}]{W09b} Wiersma R. P. C., Schaye J., Theuns T., Dalla Vecchia C., Tornatore L., 2001, MNRAS, 399, 574
\bibitem[\protect\citeauthoryear{Woo, Courteau \& Dekel}{2008}]{WCD08} Woo J., Courteau S., Dekel A., 2008, MNRAS, 390, 1453
\bibitem[\protect\citeauthoryear{Woosley \& Weaver}{1995}]{WW95} Woosley S. A., Weaver T. A., 1995, ApJS, 101, 181
\bibitem[\protect\citeauthoryear{Worthey}{1994}]{W94} Worthey G., 1994, ApJS, 95, 107
\bibitem[\protect\citeauthoryear{Yanny et al.}{2009}]{Y09} Yanny B., et al., 2009, AJ, 137, 4377
\bibitem[\protect\citeauthoryear{Yates, Kauffmann \& Guo}{2012}]{YKG12} Yates R. M., Kauffmann G., Guo Q., 2012, MNRAS, 422, 215
\bibitem[\protect\citeauthoryear{Zahid et al.}{2012a}]{Z12a} Zahid H. J., Bresolin F., Kewley L. J., Coil A. L., Dav\'{e} R., 2012a, ApJ, 750, 120
\end{thebibliography}
\end{document}